\documentclass[aps,prl,twocolumn]{revtex4-2}
\usepackage{amsmath}
\usepackage{amsfonts}
\usepackage{graphicx}

\usepackage{amssymb}
\usepackage{amsbsy}
\usepackage{color}
\usepackage{ulem}

\newcommand{\bra}[1]{\langle{#1}|}
\newcommand{\ket}[1]{|{#1}\rangle}

\newcommand{\beq}{\begin{equation}}
\newcommand{\eeq}{\end{equation}}
\newcommand{\bea}{\begin{eqnarray}}
\newcommand{\eea}{\end{eqnarray}}

\usepackage{hyperref}
\hypersetup{
	colorlinks=true,
	citecolor=blue,
	linkcolor=blue,
	urlcolor=blue}

\begin{document}

\title{ Dissipative chaos and steady state of open Tavis-Cummings dimer  }
\author{Debabrata Mondal$^1$, Andrey Kolovsky$^{2,3}$, S. Sinha$^1$}
\affiliation{$^1$Indian Institute of Science Education and
	Research-Kolkata, Mohanpur, Nadia-741246, India\\
$^2$Kirensky Institute of Physics, Federal Research Centre KSC SB RAS, 660036 Krasnoyarsk, Russia \\
$^3$School of Engineering Physics and Radio Electronics, Siberian Federal University, 660041 Krasnoyarsk, Russia	
}

\begin{abstract}
	We consider a coupled atom-photon system described by the Tavis-Cummings dimer (two coupled cavities) in an open environment, to investigate the quantum signature of dissipative chaos.
	The appropriate classical limit of this model allows us to obtain a phase diagram identifying different dynamical phases, especially the onset of chaos. 
	Both classically and quantum mechanically, we demonstrate the emergence of a steady state in the chaotic regime and analyze its properties. The interplay between quantum fluctuation and chaos leads to enhanced mixing dynamics and dephasing, resulting in the formation of an incoherent photonic fluid. The steady state exhibits an intriguing phenomenon of subsystem thermalization even outside the chaotic regime; however, the effective temperature increases with the degree of chaos. Moreover, the statistical properties of the steady state show a close connection with the random matrix theory, which, in contrast, is absent in the Liouvillian spectrum for this system. Finally, we discuss the experimental relevance of our findings, which can be tested in cavity and circuit quantum electrodynamics (QED) setups.
\end{abstract}

\maketitle

{\it Introduction: } 
Understanding the signature of chaos in quantum systems \cite{Haake_book,BGS, Izrailev_chaos} still remains a vibrant area of research over the past decade. In spite of the lack of phase-space trajectories in quantum frameworks, the Bohigas-Giannoni-Schmit (BGS) conjecture plays a pivotal role in diagnosing chaos from spectral statistics of the Hamiltonian system \cite{BGS}.
%
%
While the connection between chaos, thermalization, and ergodicity in isolated quantum setups has been extensively studied \cite{Thermalization_1,Thermalization_2,Thermalization_3, Thermalization_4, Deutsch,Srednicki}, it is less explored in open quantum systems \cite{Prosen_open_therm,Podolsky,Takashi_Mori_therm, Haake_GHS, Beneti, Casati, BHT_chaos, Deb_TCH, Vivek_OADD, Minganti, Russomanno, Ott_paper, Wimberger}, particularly regarding the fate of thermal steady states in dissipative environments \cite{Prosen_open_therm,Podolsky,Takashi_Mori_therm,Cuiti_therm_SS}.
%
In recent years there has been an impetus to study dissipative quantum chaos from spectral properties of the Liouvillian \cite{Haake_GHS,Prosen_PRX,Prosen_PRL,Prosen_JPA,Zyczkowsky,Kulkarni_Dicke,Lea_Santos_GHS}, however, such correspondence remains unclear for certain systems \cite{Lea_Santos_GHS}. 
On the other hand, the classical-quantum correspondence in collective quantum systems with an appropriate semiclassical limit can facilitate the detection of chaos in the quantum counterpart \cite{T_Brandes,Altland_Haake,Lea_Entropy,Haake_GHS,Beneti,Casati,BHT_chaos,Deb_TCH,Graham, Kolovsky_1,Kolovsky_2,Sudip_review}. 

Ultracold atomic systems coupled to cavity modes offer a pathway to study open quantum systems, where dissipation naturally arises from various loss processes \cite{Helmut_Ritsch, Esslinger_rev,Zollar_rev, Daley_dissipation,Mueller_dissipation,Hendric,Keeling_Dicke,Hemmarich,Zollar_dist_trans,Dis_tran1,Dis_tran4,Dis_tran5,Dis_tran6, Dis_tran7,Chitra,Dis_tran8,Dicke_5,Dicke_6,Dis_tran9, Dis_tran10,Dis_tran11,Spin_1_Dicke, TCD_Tomoghno}.
Such atom-photon systems have recently been realized experimentally in cavity QED setups, exhibiting intriguing nonequilibrium phenomena \cite{Serge_Haroche,Carusotto,supersolid_1,supersolid_2, Hemmarich, A_M_Rey_cavity_nature,A_M_Rey_cavity_exp, Dissipative_Rydberg,B_L_Lev,Dicke_Optica,Dicke_exp_1,Dicke_exp_2,Dicke_exp_3,JCD_EXP_2,Limit_cycle_Hemmerich}.
Within a certain regime, a collection of two-level atoms interacting with a single cavity mode can be described by the Tavis-Cummings model \cite{tc1,Eckle}, which also facilitates the study of classical-quantum correspondence due to the appropriate classical limit for a large number of atoms. Such models can also be realized in circuit QED setups \cite{Steven_Girvin_1,CQED_lattice, JCD_EXP_1}.

In this work, we explore different dynamical phases, especially the onset of  chaos and its quantum signature, in a dimer of Tavis-Cummings model under dissipation, which offers classical-quantum correspondence.
Our study focuses on the properties of the emergent steady state in the chaotic regime, particularly the combined effects of quantum fluctuations and chaos on mixing dynamics and dephasing. 
Furthermore, we delve into the issue of thermalization in this open quantum system and examine the statistical properties of the steady-state density matrix. In contrast, spectral correlation in the Liouvillian is absent in the chaotic regime.

{\it Model and semiclassical analysis:} 
We consider two coupled cavities, each containing $N$ two-level atoms interacting with a single cavity mode, which can be described by the Tavis-Cummings dimer (TCD) model with the following Hamiltonian
\begin{eqnarray}
		\hat{\mathcal{H}} &=&-J \left( \hat{a}_{\mathrm L}^\dagger \hat{a}_{\mathrm R} + \hat{a}_{\mathrm R}^\dagger \hat{a}_{\mathrm L}\right)\!+\!\!\sum_{i=\rm L, R} \left[\omega \hat{a}_i^\dagger\hat{a}_i + \omega_0 \hat{S}_{zi}\right. \nonumber \\
		&&\!\!\!+ \left.\frac{\lambda}{\sqrt{2S}}\left(\hat{a}_i^\dagger \hat{S}_{i}^- + \hat{a}_i \hat{S}_{i}^+\right) 
		 \!\right],
		\label{ADM}
\end{eqnarray}
where the site index $i=\rm L(R)$ represents left(right) cavity, $\hat{a}_{i}$ annihilates photon mode with frequency $\omega$, and 
$J$ is the hopping amplitude of the photons between the cavities.
Collectively, $N$ two-level atoms with energy gap $\omega_0$ can be represented by large spins $\hat{\vec{S_{i}}}$ with magnitude $S =N/2$, and $\lambda$ is the atom-photon coupling strength.
Note that the total excitation $\hat{\mathcal{N}} =  \sum_i
(\hat{a}_i^{\dagger}\hat{a}_i+S+\hat{S}_{zi})$ is conserved as a result of ${\rm U}(1)$ symmetry of $\hat{{\mathcal H}}$ \cite{Eckle}. 
The integrability of the single-cavity Tavis-Cummings model is broken by coupling two such cavities.
Note that, a cavity containing a single atom ($S=1/2$) is reduced to the integrable Jaynes-Cummings model \cite{JC_model} and the dimer of two such coupled cavities has already been studied both theoretically \cite{JCD_1,JCD_2,JCD_3} and experimentally in cavity and circuit QED setups \cite{JCD_EXP_1,JCD_EXP_2}.  

%
Photon loss from cavities can be compensated by atomic pumping, leading to the non-unitary evolution of the density matrix  (DM) $\hat{\rho}$ described by the Lindblad master equation \cite{GKS,Lindblad,Breuer},
\begin{small}
\begin{eqnarray}
\dot{\hat{\rho}} = -i[\mathcal{\hat{H}},\hat{\rho}]+ \kappa \sum_i\mathcal{D}[\hat{a}_i] +\frac{1}{S}\sum_i(\gamma_{\uparrow} \mathcal{D}[\hat{S}_{i}^+]+
\gamma_{\downarrow}
\mathcal{D}[\hat{S}_{i}^-]),\,\,\,
	\label{dmeq1}
\end{eqnarray}
\end{small}
with $\mathcal{D}[\hat{L}] = \frac{1}{2} \left(2\hat{L}\hat{\rho} \hat{L}^{\dagger}-\hat{L}^{\dagger}\hat{L}\hat{\rho}-\hat{\rho} \hat{L}^{\dagger}\hat{L}\right)$ describing the dissipative process corresponding to the Lindblad operator $\hat{L}$. 
The dissipators $\hat{L}=\hat{a}_i$ and $\hat{S}_{i}^-$ account for the decay processes of photon and spins with amplitudes $\kappa$ and $\gamma_{\downarrow}$ respectively.
The incoherent pumping of the atoms is represented by $\mathcal{D}[\hat{S}_{i}^+]$ with rate $\gamma_{\uparrow}>\gamma_{\downarrow}$.
From the time evolved DM, we obtain an average of any operator $\hat{O}$ using $\langle \hat{O}\rangle={\rm Tr}(\hat{\rho}\hat{O})$. 
Note that, the excitation number $\hat{\mathcal{N}}$ is no longer conserved in the presence of dissipation.
Throughout the paper, we set $\hbar,k_B=1$ and scale energy (time) by $J$ $(1/J)$.

For $S\gg 1$, the scaled operators $\hat{\alpha}_i = \hat{a}_{i}/\sqrt{S}$ and $\hat{\vec{s}}_{i} = \hat{\vec{S}}_{i}/S$ become classical, as they satisfy $[\hat{\alpha}_{i},\hat{\alpha}_{i}^{\dagger}]= 1/S$ and $[\hat{s}_{ai}, \hat{s}_{b i}] = \imath \epsilon_{abc} \hat{s}_{c i}/S$,
where $1/S$ plays the role of reduced Planck constant. Within the mean-field approximation, the expectation of the product of operators can be decomposed as $\langle \hat{A} \hat{B}\rangle = \langle \hat{A} \rangle \langle \hat{B} \rangle$, which is valid for $ S \rightarrow \infty$ \cite{Sudip_review,Dis_tran11, Dicke_correlator}.  
The semiclassical dynamics of the scaled observables is obtained from Eq.\eqref{dmeq1}, which is described by the following equations of motion (EOM),
\begin{subequations}
	\begin{eqnarray}
		\dot{\alpha_i} &=& -\left(\kappa/2+\imath\omega\right) \alpha_i-\frac{\imath}{\scalebox{0.8}{$\sqrt{2}$}}\lambda s_{i}^-+\imath \alpha_{j} \\
		\dot{s}_{i}^+ &=& \imath\omega_0 s_{i}^+-\imath \lambda \scalebox{0.8}{$\sqrt{2}$} s_{zi}\alpha_i^*-f_c s_{zi}s_i^+\\
		\dot{s}_{zi} &=& -\frac{\imath}{ \scalebox{0.8}{$\sqrt{2}$} } \lambda(\alpha_i s_{i}^+-\alpha_i^*s_{i}^-)+f_c(1-s_{zi}^{2}),
	\end{eqnarray}
	\label{EOM}
\end{subequations}
where  $j\ne i$, $f_c=\gamma_{\uparrow}-\gamma_{\downarrow}$, $\alpha_i = \sqrt{n_i}\exp(-i\psi_i)=(x_i+\iota p_i)/\sqrt{2}$ and $\vec{s}_i = (\sin\theta_i\cos\phi_i,\sin\theta_i\sin\phi_i,\cos\theta_i)$. 

\begin{figure}[h]
	\includegraphics[width=\linewidth]{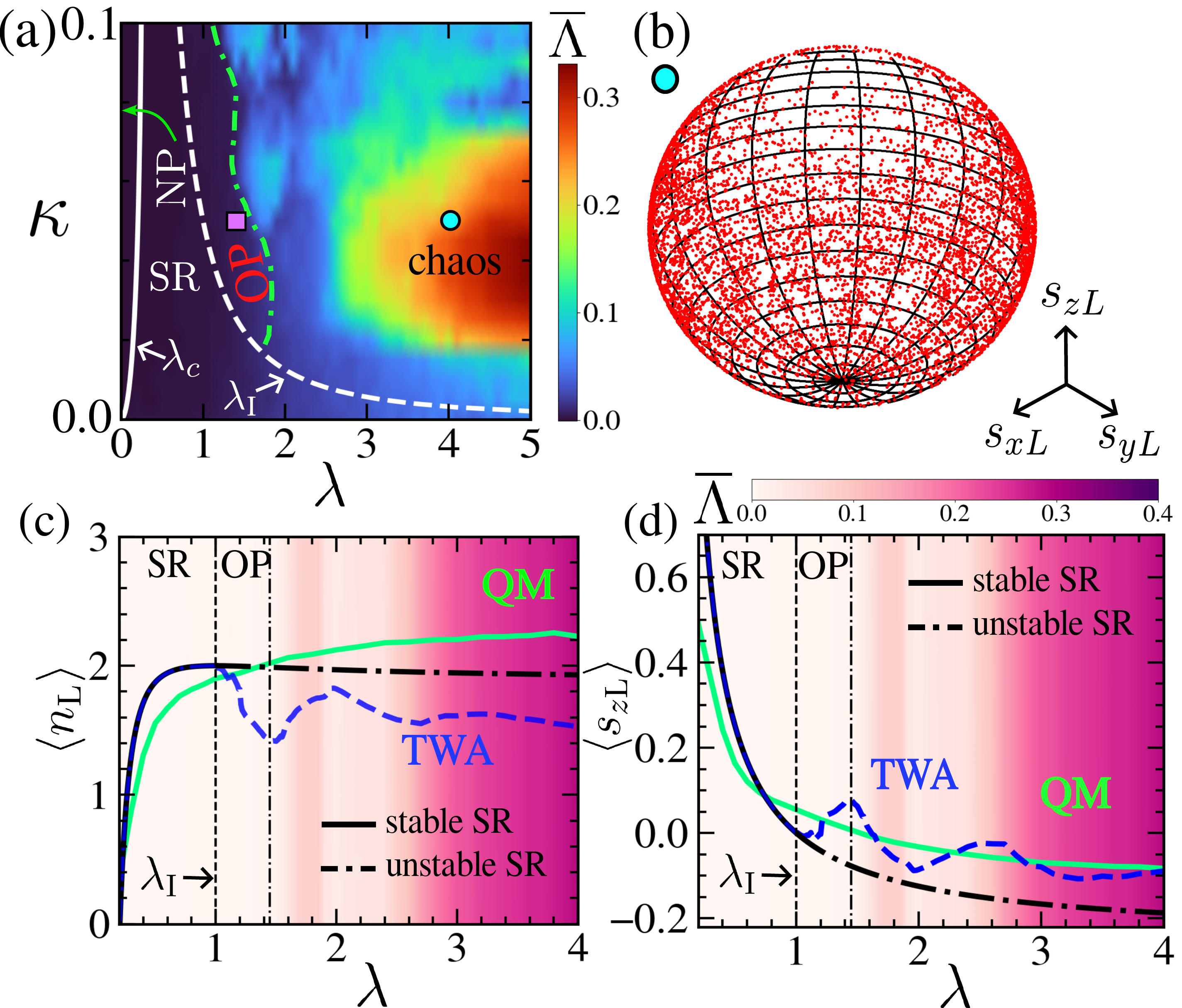}
	\caption{{\it Dynamical phases:} (a) Classical phase-diagram in $\lambda-\kappa$ plane with average Lyapunov exponent $\bar{\Lambda}$ as color scale. 
	The NP to SR transition at $\lambda_c$ (solid line) and instability of SR phase at $\lambda_{\rm I}$ (dashed line) are shown. 
	(b) Chaotic trajectory is represented by scattered points on the Bloch sphere at regular time intervals ($\lambda = 4.0,\kappa=0.05$ (blue circle in (a))).
	{\it Steady state}: Variation of average (c) photon number $\langle n_{\rm L}\rangle$ and (d) spin polarization $\langle s_{z{\rm L}}\rangle$ of one cavity with $\lambda$ for $\kappa=0.05$. TWA results (blue dashed line), quantum steady-state (solid green line) and fixed point of SR phase are compared. 
	Stable (unstable) SR phase is shown by black solid (dashed dotted) line.
	The color scales in (c,d) represent $\bar{\Lambda}$. All energies (time) are measured by $J (1/J)$. We set $\hbar,k_B = 1$ and $\omega=2.0,\omega_0=0.5, \gamma_{\uparrow}=0.2,\gamma_{\downarrow}=0.1, S=2$ for all figures. }
	\label{fig1}
\end{figure}

Next, we investigate the stable fixed points (FP) and other attractors of the EOM (see Eq.\eqref{EOM}), describing various nonequilibrium phases of this open TCD, listed below and summarized in the phase diagram in Fig.\ref{fig1}(a), which is plotted in the $\lambda$-$\kappa$ plane for a fixed spin pumping rate $f_c$.

\textbf{Normal phase (NP)}: This phase is characterized by vanishing photon number $n_i^*=0$ and spin polarization $s_{zi}^*=1$.

\textbf{Superradiant phase (SR)}: At a critical coupling $\lambda_c$ (depending on the dissipation strengths $\kappa$ and $f_c$), the normal phase undergoes a continuous transition to the superradiant phase with a non-vanishing photon number $n^*\ne0$ and spin polarization $|s_{z}^*|<1$, same for each cavity. 
%
%
In this case, the incoherent pumping of spins balances the photon loss, resulting in the formation of SR phase.

Due to the U(1) symmetry, FPs lie on a circle of radius $\sqrt{2n^*}$ ($\sqrt{1-(s_{z}^*)^2}$) in $x-p$ ($s_{x}-s_{y}$) plane, where the dynamics always converges regardless of the initial condition. 
The details of this phase are given in the supplementary material \cite{SM}.

\textbf{Oscillatory phase (OP):} 
Once the SR phase becomes unstable for  $\lambda>\lambda_{\rm I}$ (see Fig.\ref{fig1}(a,c,d)) \cite{foot_note_0}, the oscillatory phase emerges, where the periodic motion is identified from a single peak in the Fourier transform of trajectories \cite{SM}. Although the photon number oscillates, its phase in both cavities remains the same $\psi_{\rm L}=\psi_{\rm R}$, similar to the SR phase.

\textbf{Coexistance of chaos and oscillatory dynamics:} Further increasing $\lambda$ gives rise to the mixed type of dynamics, where oscillatory motion coexists with chaotic dynamics, depending on the initial conditions.

\textbf{Chaotic dynamics:} We also identify a regime in the phase diagram where the trajectories exhibit chaotic behavior, as depicted in Fig.\ref{fig1}(b).
To quantify the degree of chaos, we compute the mean Lyapunov exponent $\bar{\Lambda}$ \cite{Strogatz, Lichtenberg,Lyapunov} averaged over random initial phase space points, as shown in Fig.\ref{fig1}(a). The chaos is suppressed by the appearance of stable fixed points such as SR phase.

\begin{figure}[t]
	\includegraphics[width=\linewidth]{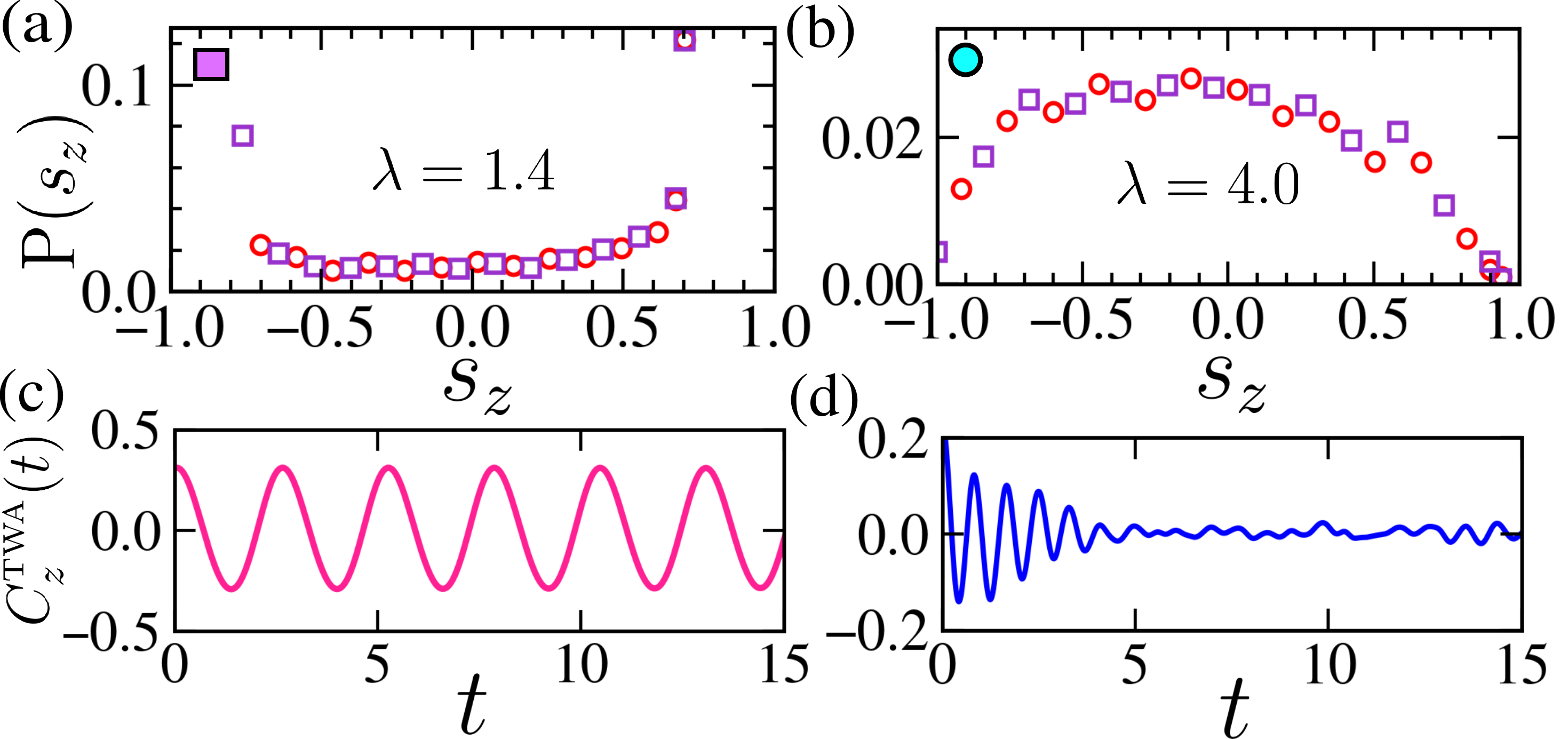}
	\caption{{\it Comparison between oscillatory and chaotic phases:}  Steady-state distribution and autocorrelation function of $s_{z}$ of one cavity in the (a,c) oscillatory ($\lambda = 1.4$, violet square in Fig.\ref{fig1}(a)), (b,d) chaotic regime ($\lambda = 4.0$, blue circle in Fig.\ref{fig1}(a)). Squares and circles in (a,b) represent distributions from different initial conditions. We choose $\kappa=0.05$. 
	}
	\label{fig2}
\end{figure}

The stable fixed point attractors, such as NP and SR phases, uniquely describe the system's asymptotic steady state. To understand the state of the system after the instability of the SR phase, we study the semiclassical dynamics 
using truncated Wigner approximation (TWA) \cite{TWA_1,TWA_2,A_M_Rey_TWA}.
To mimic quantum fluctuations, we perform TWA by sampling initial conditions from the Husimi distribution \cite{Q_traj_3} of product states of bosonic and spin coherent states, $\ket{\Psi_c}=\prod_i|\alpha_i \rangle \otimes |s_{zi},\phi_i \rangle$ \cite{Coherent_state} corresponding to large spin $S\gg 1$, which semiclassically represents an arbitrary phase space point $\{\alpha_i, s_{zi}=\cos\theta_i,\phi_i\}$.
Importantly, we observe the emergence of a unique steady state in both oscillatory and chaotic phases, characterized by the stationary value of physical quantities such as photon number $\langle n \rangle_{\scriptscriptstyle\text{TWA}}$ and spin polarization $\langle s_{z} \rangle_{\scriptscriptstyle\text{TWA}}$, which smoothly connects to the stable SR phase, as shown in Fig.\ref{fig1}(c,d).
Moreover, these dynamical variables follow stationary distribution P$(n_i)$, P$(s_{zi})$ irrespective of initial condition (see Fig.\ref{fig2}(a,b)).
%
%
Notably, our analysis reveals the ergodic nature of the steady state, which is based on (i) the equivalence between time and ensemble averages $\langle O\rangle_t = \langle O\rangle_{\scriptscriptstyle\text{TWA}}$ of dynamical quantities and (ii) their independence from the initial condition \cite{Sinai_2,Halmos,Ott,Berkovitz}. 

The different dynamical regimes of the steady state particularly, the onset of chaos can be revealed from the autocorrelation function, 
\begin{eqnarray}
	C^{\scriptscriptstyle\text{TWA}}(t)=\langle O(t+\tau)O(\tau)\rangle-\langle O(t+\tau)\rangle \langle O(\tau)\rangle,
	\label{Autocorrelation}
\end{eqnarray}
computed within TWA. Here, the observable $O$ is first evolved up to the time $\tau$ until $\langle O(\tau)\rangle_{\scriptscriptstyle\text{TWA}}$ reaches the steady state.
Although the TWA analysis indicates the formation of a steady state in the oscillatory regime, the autocorrelation function exhibits persistent oscillations, revealing its signature (see Fig.\ref{fig2}(c)). On the contrary, in the chaotic regime, the autocorrelation function decays rapidly, as seen from Fig.\ref{fig2}(d), confirming chaotic mixing \cite{Berkovitz,Ruelle,Ott}. 
Interestingly, the steady state in the oscillatory regime displays ergodicity in the absence of mixing. This contrasts with the typical closed system, where chaotic mixing leads to the ergodic steady state.

{\it Quantum steady state:} To this end, we investigate the properties of the quantum steady state and the signature of chaos in presence of quantum fluctuation. 
We obtain the steady-state density matrix (DM) $\hat{\rho}_{\rm ss}$ by solving the Master equation (see Eq.\eqref{dmeq1}) using the stochastic wavefunction approach \cite{Q_traj_1,Q_traj_2,Q_traj_3}, considering spin magnitude $S=2$ and sufficient number of photonic states for each cavity to ensure negligible truncation error.
Although larger values of $S$ are better for classical correspondence, it is computationally more expensive.
%
\begin{figure}[b]
	\includegraphics[width=\linewidth]{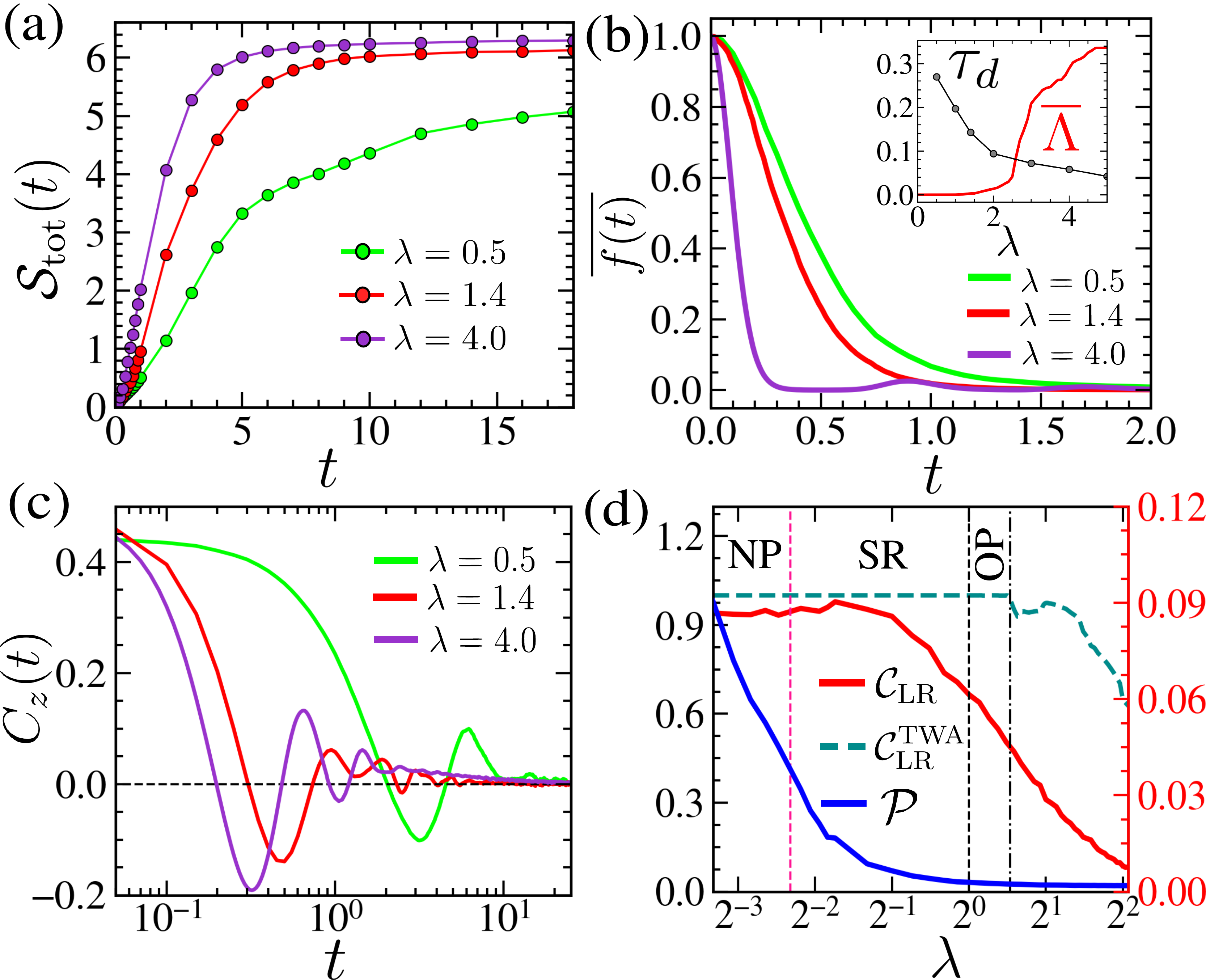}
	\caption{{\it Quantum dynamics:} (a) Evolution of total entropy $\mathcal{S}_{\rm tot}$ and (b) survival probability $\overline{f(t)}$ at different $\lambda$. Inset in (b) shows decay time $\tau_d$ of $\overline{f(t)}$ (black line) and average Lyapunov exponent $\overline{\Lambda}$ (red line) as a function of $\lambda$. (c) Dynamics of spin autocorrelation $C_z(t)$ for different $\lambda$. (d) Coherence function $\mathcal{C}_{\rm LR}$ of the photon field between the cavities (solid red), corresponding TWA result (dashed green) and the purity $\mathcal{P}$ (solid blue, right axis), as a function of $\lambda$. Parameter: $\kappa=0.05$. }
	\label{fig3}
\end{figure}
%
Similar to the classical analysis, a unique quantum steady state is formed across different dynamical regimes with increasing coupling strength. This state is characterized by the average quantities $\langle \hat{n}_i \rangle_{\rm ss}$ and $\langle \hat{S}_{zi} \rangle_{\rm ss}$ obtained  from $\hat{\rho}_{\rm ss}$, as shown in Fig.\ref{fig1}(c,d). For comparison with classical results, we scale these physical quantities by $S$.
The steady state also exhibits ergodicity analogous to its classical counterpart, as the time average of the observables $\langle \hat{n}_i \rangle_t$ and $\langle \hat{S}_{zi} \rangle_t$ over a typical quantum trajectory approach their steady-state values $\langle \hat{n}_i \rangle_{\rm ss}$ and $\langle \hat{S}_{zi} \rangle_{\rm ss}$ respectively \cite{SM}.

Next, we study the nonequilibrium dynamics to investigate the signature of quantum mixing, as the chaotic regime is approached with increasing $\lambda$.
Starting from an arbitrary initial coherent state $\ket{\Psi_c}$, representing a phase space point, we study the time evolution of survival probability $f(t)= {\rm Tr}(\hat{\rho}(t)\hat{\rho}(0))$ \cite{Nilson_Chuang,Carmichael_fidelity,Fazio_fidelity,footnote} between the initial and time-evolved density matrices.
%
The survival probability $\overline{f(t)}$, averaged over an ensemble of initial states, decays exponentially with time and its asymptotic decay rate can be influenced by the spectral gap (the largest real part of the eigenvalues of the Liouvillian) \cite{DSFF_Circuit}. 
In Fig.\ref{fig3}(b), we investigate $\overline{f(t)}$ for increasing values of the coupling strength $\lambda$ while keeping 
$\kappa$ fixed, to eliminate the effect of dissipation strength on the variation of the decay rate.
We observe that the decay rate increases with an enhanced degree of chaos (quantified by the Lyapunov exponent) as the coupling strength $\lambda$ increases (see Fig.\ref{fig3}(b)). 
The connection between quantum chaos and the relaxation process has also been reported in other model \cite{DSFF_Circuit}.

\begin{figure}
	\includegraphics[width=\linewidth]{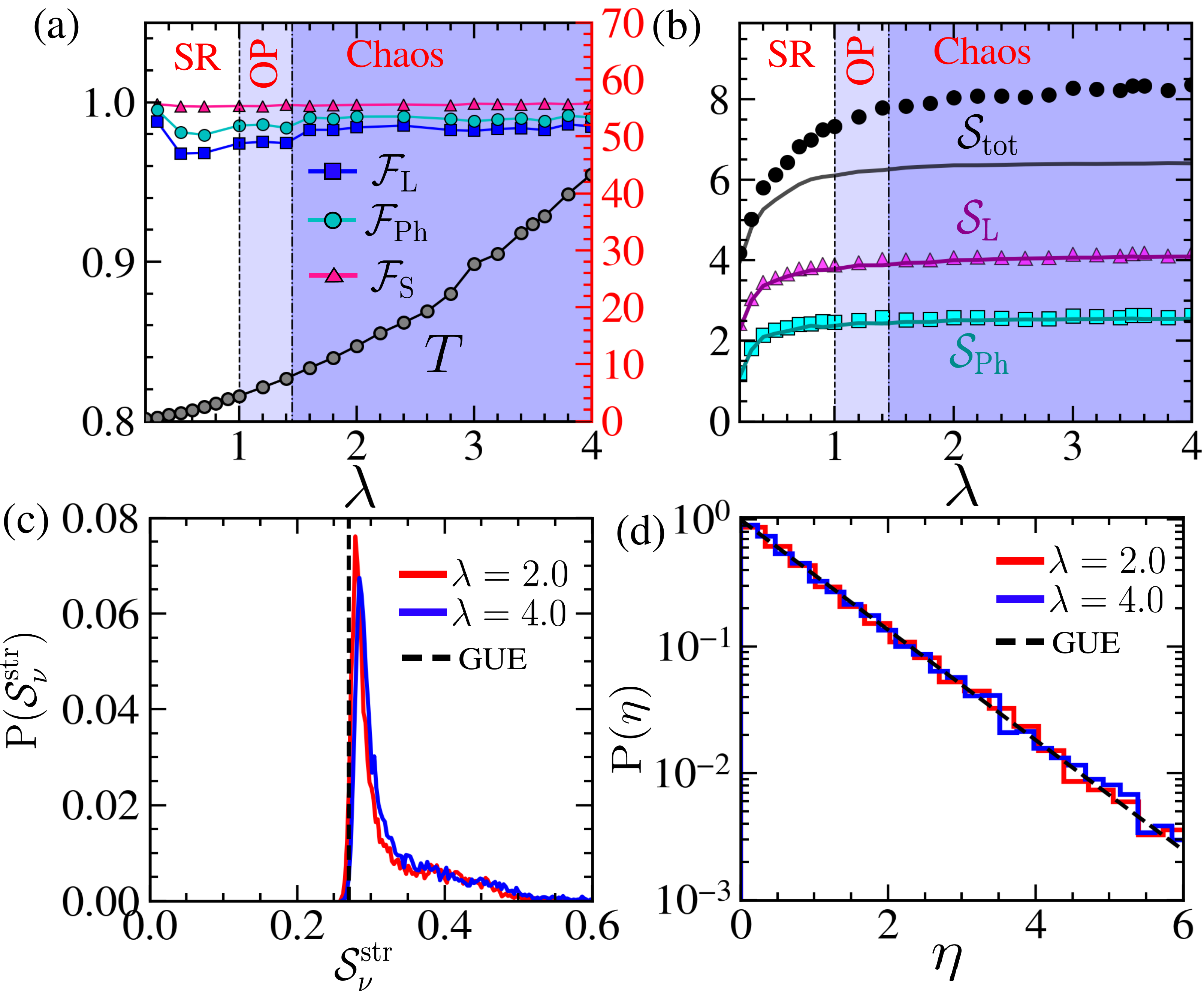}
	\caption{{\it Subsystem thermalization:} (a) Overlap $\mathcal{F}_{A}$ between reduced DMs of $\hat{\rho}_{\rm ss}$ and the thermal state $\hat{\rho}^{\rm th}$. Overlap for the left cavity ($\mathcal{F}_{\rm L}$), photon ($\mathcal{F}_{\rm Ph}$), spin ($\mathcal{F}_{\rm S}$) of the same cavity (left axis) and effective temperature $T$ (right axis) as a function of $\lambda$. (b) Total entropy $\mathcal{S}_{\rm tot}$, entanglement entropy of the left cavity $\mathcal{S}_{\rm L}$ and photon field of that cavity $\mathcal{S}_{\rm Ph}$ in the steady state $\hat{\rho}_{\rm ss}$ versus $\lambda$. Solid lines (markers) represent the entropies obtained from $\hat{\rho}_{\rm ss}$ ($\hat{\rho}^{\rm th}$).  Classical phases are marked by blue-colored regimes. {\it Statistics of $\hat{\rho}_{\rm ss}$:} (c) distribution of structural entropy $\mathcal{S}_{\nu}^{\rm str}$ of eigenstates of $\hat{\rho}_{\rm ss}$ in the chaotic regime. (d) Distribution of elements $\eta=|\Psi_{\nu}^j|^2N_{\rm dim}$ of typical eigenstate $\ket{\Psi_{\nu}}$. Black dashed lines in (c,d) represent the corresponding GUE results. Parameter chosen: $\kappa=0.05$.  }
	\label{fig4}
\end{figure}

From the time evolved density matrix, we also obtain the total entropy $\mathcal{S}_{\rm tot}=-{\rm Tr}(\hat{\rho}(t)\ln\hat{\rho}(t))$, which grows linearly and attains a saturation value corresponding to the steady state.
As evident from Fig.\ref{fig3}(a), both the growth rate and the saturation value of $\mathcal{S}_{\rm tot}$ increase with $\lambda$. Moreover, the evolution of the entanglement entropy (EE) $\mathcal{S}_A=-{\rm Tr}(\hat{\rho}_A\ln\hat{\rho}_A)$ of subsystem $A$ obtained from the reduced DM  $\hat{\rho}_A={\rm Tr}_{\bar{A}}(\hat{\rho})$ by tracing out the remaining degrees of freedom $\bar{A}$, also exhibits similar behavior with coupling \cite{SM}.
Such linear growth in EE is typically observed as a signature of chaos in isolated quantum systems \cite{Thermalization_1,Thermalization_3, EE_chaos_1,EE_chaos_2, EE_chaos_3, EE_chaos_4}. 
However, in certain integrable systems, special initial states can lead to similar entropy production rate \cite{Int_entropy}.
The underlying chaos can also be unveiled from the individual quantum trajectories, which show a spreading of the power spectrum over a wide range of frequencies \cite{SM,T_Geisel}.

We also analyze the autocorrelation function corresponding to the steady state $C(t)=\langle \hat{O}(t)\hat{O}(0)\rangle-\langle \hat{O}(t)\rangle \langle \hat{O}(0)\rangle$, following the prescription of stochastic wave-function \cite{Correlation_1,Q_traj_2}.
As seen from Fig.\ref{fig3}(c), the autocorrelation function $C_z(t)$ of the spin in one cavity vanishes rapidly and the decay time decreases with increasing coupling strength $\lambda$.
Such fast decay of $C_z(t)$ signifies enhanced mixing dynamics due to the onset of chaos. 

Additionally, quantum fluctuations enhance the decay rate of $C_z(t)$ compared to the classical counterpart. 
Importantly, the oscillatory phase is washed out due to the strong quantum fluctuation, which is evident from the decaying autocorrelation function (see Fig.\ref{fig3}(c)).
Furthermore, the combined effect of quantum fluctuations and chaotic mixing suppresses the purity of the steady state $\mathcal{P}={\rm Tr}(\hat{\rho}_{\rm ss}^2)$ significantly, even in the superradiant phase, depicted in Fig.\ref{fig3}(d). 
The coherence of the photon fields between two cavities in the steady state can be quantified from,
\begin{eqnarray}
	\mathcal{C}_{\rm LR} = \frac{\langle \hat{a}_{\rm L}^{\dagger}\hat{a}_{\rm R}+\hat{a}_{\rm R}^{\dagger}\hat{a}_L\rangle}{2\sqrt{\langle \hat{n}_{\rm L}\rangle\langle \hat{n}_{\rm R}\rangle }},
\end{eqnarray}
which classically reduces to $\mathcal{C}_{\rm LR}^{\rm TWA}=\langle \cos(\psi_{\rm L}-\psi_{\rm R})\rangle_{\scriptscriptstyle\text{TWA}}$. In the superradiant and oscillatory phase, $\mathcal{C}_{\rm LR}^{\rm TWA}$ attains the value unity and decays in the chaotic regime, as observed from the TWA analysis. Similar behavior can also be observed quantum mechanically, however, the decay starts even before the instability of the SR phase, as a result of quantum fluctuations (see Fig.\ref{fig3}(d)). It is clear from this analysis that the mixing dynamics due to the underlying chaos as well as quantum fluctuation can destroy the coherence of the system, which can be probed from the state of the photon field of the respective cavities. 
The semiclassical distribution of the photon field in the stable SR phase for small value of $\lambda$ is peaked around the circle of classical fixed points of radius $\sqrt{2n^*}$. 
As we approach the chaotic regime, this distribution spreads and peaks around the center, resembling a thermal distribution \cite{SM}. Such an observation suggests chaos-induced thermalization in this dissipative system, which we discuss next.

{\it Subsystem thermalization:} To investigate thermalization of this open system, we consider the thermal density matrix of TCD $\hat{\rho}^{\rm th}=\exp(-\beta(\hat{\mathcal{H}}-\mu\hat{\mathcal{N}} ))/Z$ with $Z$ being the partition function. The inverse temperature $\beta$ and the chemical potential $\mu$ are uniquely determined from the mean energy $\langle \hat{\mathcal{H}}\rangle$ and the average number of excitations $\langle \hat{\mathcal{N}}\rangle$ obtained from the steady-state density matrix $\hat{\rho}_{\rm ss}$. 
Apart from certain systems with special choice of dissipators \cite{Thermal_Steady_state}, in general, the steady state of an open system as a whole does not correspond to a thermal state \cite{Prosen_open_therm}.
In the present case, clearly, $\hat{\rho}^{\rm th}$ does not correspond to the steady state of the Master equation (Eq.\eqref{dmeq1}) in the presence of dissipation. Consequently, the total entropy of the thermal state $\hat{\rho}^{\rm th}$ is significantly deviated from that of the steady state $\hat{\rho}_{\rm ss}$, as shown in Fig.\ref{fig4}(b).
However, the state of the subsystem $A$ can be well described by the reduced DM $\hat{\rho}_A^{\rm th}$ derived from the thermal state $\hat{\rho}^{\rm th}$ of the full system. We compare the reduced DMs $\hat{\rho}_A^{\rm ss}$ and $\hat{\rho}_A^{\rm th}$ obtained from the steady state and thermal DM respectively, using their overlap $\mathcal{F}_A = {\rm Tr}\sqrt{\sqrt{\hat{\rho}_A^{\rm ss}}\hat{\rho}_A^{\rm th}\sqrt{\hat{\rho}_A^{\rm ss}}}$ \cite{Nilson_Chuang,Carmichael_fidelity}. Considering one of the cavities as well as its photonic and spin sector as subsystems, we compute $\mathcal{F}_A$, which remains close to unity, as evident from Fig.\ref{fig4}(a). Moreover, trace-distances between  $\hat{\rho}_A^{\rm ss}$ and $\hat{\rho}_A^{\rm th}$ remain small \cite{SM},  indicating the similarity between them.
Consequently, the entanglement entropies of the corresponding subsystems (see Fig.\ref{fig4}(b)) as well as the average values of the observables such as $\langle \hat{n}_i\rangle$ and $\langle \hat{S}_{zi}\rangle$ exhibit excellent agreement with those of the thermal state \cite{SM}, confirming the validity of subsystem thermalization \cite{Sub_therm_1,Sub_therm_2,Sub_therm_3} in open TCD. 
Surprisingly, the steady state follows the subsystem thermalization over a range of $\lambda$ even outside the chaotic regime, however, the effective temperature $T$ increases with the degree of chaos, as shown in Fig.\ref{fig4}(a).

For a deeper understanding of this scenario, we also study the statistics of eigenstates $\ket{\Psi_{\nu}}=\sum_{j=1}^{N_{\rm dim}}\Psi_{\nu}^j\ket{j}$  of the steady-state DM $\hat{\rho}_{\rm ss}$, where $\ket{j}$ are the computational basis states. We compute the structural entropy \cite{RMT_GUE_1,RMT_GUE_2} of $\ket{\Psi_{\nu}}$,
\begin{eqnarray}
	\mathcal{S}_{\nu}^{\rm str}= -\sum_{j}|\Psi_{\nu}^j|^2\ln(|\Psi_{\nu}^j|^2)+\ln\left(\sum_{j}|\Psi_{\nu}^j|^4\right),
\end{eqnarray}
and study its distribution, which is sharply peaked around a value $\mathcal{S}^{\rm str}\approx 0.27$ (see Fig.\ref{fig4}(c)) corresponding to the Gaussian unitary ensemble (GUE) class of the random matrix theory (RMT) \cite{Izrailev_chaos,Bhaskar}. Moreover, the elements $\eta = |\Psi_{\nu}^j|^2N_{\rm dim}$ of the eigenstate $\ket{\Psi_{\nu}}$ (with dimension $N_{\rm dim}$), follows the GUE distribution  ${\rm P}(\eta)=\exp(-\eta)$ \cite{Haake_book}, as seen from Fig.\ref{fig4}(d), indicating the chaotic nature of such states. 
The steady state of a random Liouvillian also follows the GUE distribution \cite{Prosen_JPA,Prosen_GUE}. However, in the present case, we do not find any signature of level repulsion in the Liouvillian spectrum, which on the contrary resembles to 2d-Poisson distribution, in the chaotic regime. On the other hand, it exhibits Ginibre correlation as the chaos is suppressed by increasing the dissipation strength $\kappa$ \cite{SM}, indicating a violation of GHS conjecture \cite{Haake_GHS, Lea_Santos_GHS}. Due to the complexity of the present system, the spectral statistics of the Liouvillian is analyzed with a small photon-number truncation, constrained by computational capacity. Additionally, achieving better classical-quantum correspondence would require larger spin magnitudes $S$, which is also computationally demanding. 

{\it Conclusion:} 
%
Our study unveils the emergence of a steady state with intriguing properties,  linked to the onset of chaos in an open atom-photon dimer system. 
%
Unlike a generic quantum system, the present setup allows us to explore dissipative chaos using classical-quantum correspondence.
%
%
%
The rapid decay of the correlation function and survival probability, along with the growth of entropy, indicating the chaotic mixing, serve as tangible signatures of dissipative chaos.
As the decay of survival probability in an open system can depend on dissipation, we identify fast mixing due to the onset of chaos from its increasing decay rate with coupling $\lambda$, while keeping $\kappa$ fixed to eliminate its dependence on dissipation. A larger spin magnitude $S$ is preferable for better classical-quantum correspondence, as enhanced quantum fluctuations for small $S$ can also contribute to fast mixing, even outside the classically chaotic regime.

Both chaos and quantum fluctuations result in the loss of coherence, leading to the formation of an incoherent photonic fluid, which can be probed experimentally. 
Although, except for specific systems, the steady state of an open quantum system generally does not correspond to a thermal state, remarkably, the steady state in the present case exhibits subsystem thermalization and reveals a connection to random matrix theory.
The dissipative steady state contains more detectable signatures of chaos than Liouvillian statistics.

To summarize, this atom-photon dimer system exhibits intriguing nonequilibrium phenomena and sheds light on dissipative quantum chaos, with results that are readily testable in current cavity and circuit QED setups.

\begin{acknowledgments}
{\it Acknowledgments:} We thank Krishnendu Sengupta and Sudip Sinha for comments and fruitful discussions. A.K.  acknowledges supports from Ministry of Science and Higher Education of Russian Federation, Grant No. FSRZ-2023-0006. D.M. acknowledges support from Prime Minister Research Fellowship (PMRF).

\end{acknowledgments}


\begin{thebibliography}{99}
\bibitem{footnote}
For unitary dynamics of initial pure state $\hat{\rho}(0)=\ket{\Psi(0)}\bra{\Psi(0)}$, the quantity $f(t)= {\rm Tr}(\hat{\rho}(t)\hat{\rho}(0))$ reduces to the usual definition of survival probability $f(t)= |\langle \Psi(t)|\Psi(0)\rangle|^2$. Additionally, it is related with the more general definition of overlap between two mixed density matrices (see Ref. \cite{Nilson_Chuang,Carmichael_fidelity}).
\end{thebibliography}
\end{document}


\begin{center}
	{\bf SUPPLEMENTARY MATERIAL:\\
		Dissipative chaos and steady state of open Tavis-Cummings dimer   }
\end{center}

\section{Classical analysis}
For the coupled atom-photon system, a phase-space point can be represented by an array $\mathbf{X}=\{\alpha_i,s_{zi}=\cos\theta_i,\phi_i\}$, where both the cavities $i={\rm L,R}$ are included. The fixed points (FP) $\mathbf{X}^*$ representing the steady states, satisfying $\dot{\mathbf{X}}=0$, can be obtained from the equations of motion (given in Eq.(3) of the main text). To investigate the stability of the different phases of the open Tavis-Cummings dimer (TCD), we perform the linear stability analysis.
We write the dynamical variables as $\mathbf{X}(t) = \mathbf{X}^* +\delta \mathbf{X}(t)$, representing the small fluctuation $\delta \mathbf{X}(t)$ around the FP $\mathbf{X}^*$. The time evolution of the fluctuations can  be written as $\delta\mathbf{X}(t) = \delta \mathbf{X}(0) e^{\lambda_s t}$, where $\lambda_s$ are obtained from the linearized equations and the stability of a steady state is ensured by the condition ${\rm Re}(\lambda_s)<0$ \cite{Strogatz}.
%
%
Following this procedure, we have obtained the stability regimes of the various steady states in the parameter space, as given in the phase diagram (see Fig.1(a) of the main text for details). 
\begin{figure}[h]
	\includegraphics[width=0.8\linewidth]{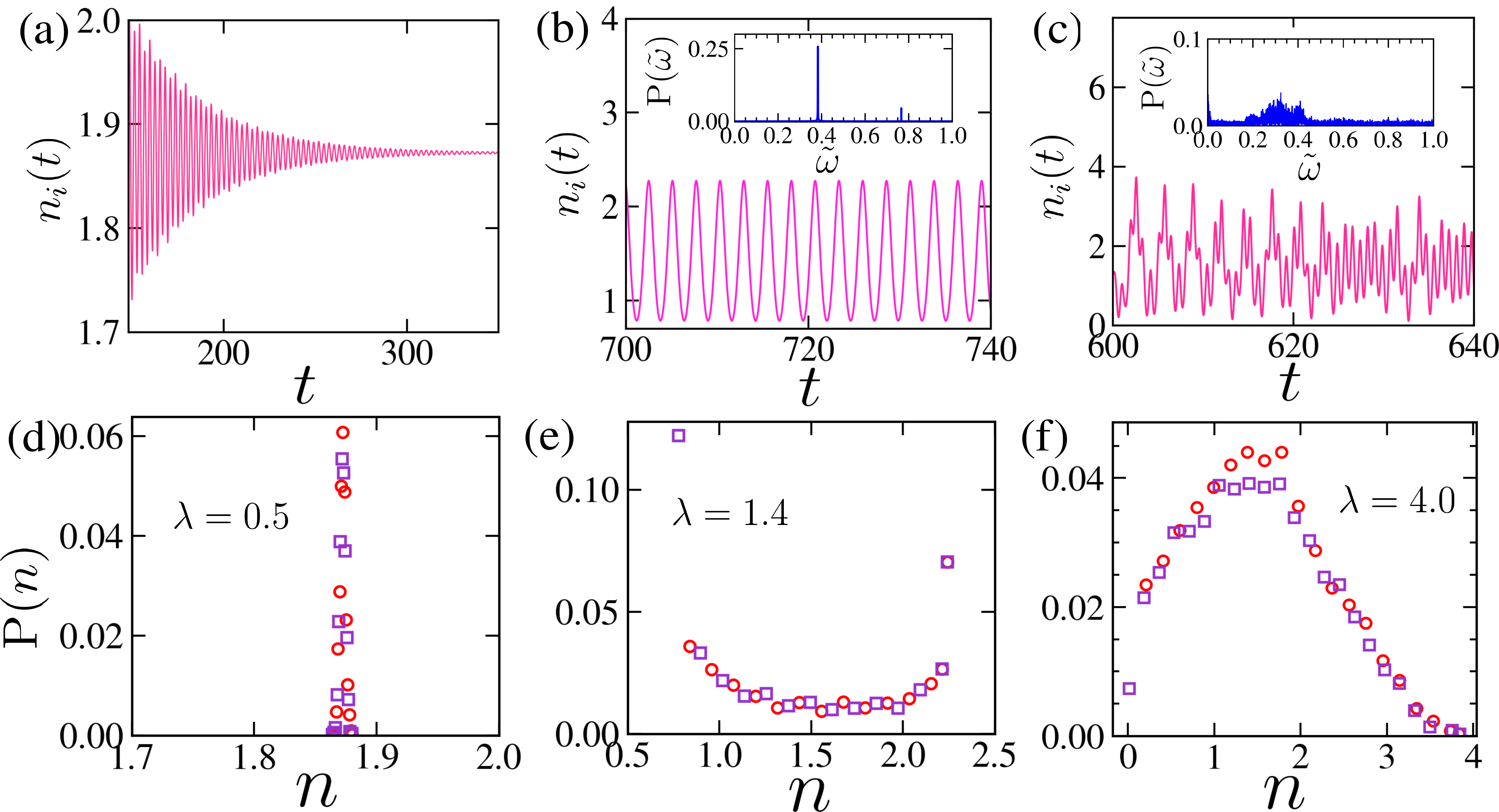}
	\caption{ {\it Classical dynamics:} Time evolution of the photon number $n_i(t)$ of one of the cavities in the (a) SR phase ($\lambda=0.5$), (b) oscillatory phase ($\lambda=1.4$) and (c) chaotic regime ($\lambda=4.0$). The inset of (b), (c) show the Fourier spectrum of the corresponding trajectory.
	Distribution of the photon number $n_i$ at long time within TWA for the (d) SR state ($\lambda = 0.5$), (e) oscillatory phase (OP) ($\lambda = 1.4$) and the (f) chaotic phase ($\lambda = 4.0$). Squares and circles in (d-f) represent distributions starting from different initial conditions.
	We consider $\omega=2.0,\omega_0=0.5,\kappa = 0.05, f_c = 0.1$. }
	\label{fig1}
\end{figure}

We summarize the homogenous phases ($\mathbf{X}_i = \mathbf{X}^*$) of open TCD and transitions between them. The normal phase (NP) is characterized by vanishing photon number $n^*=0$ and spin polarization $s_{z}^*=1$. The NP becomes unstable at a critical coupling $\lambda_c$ and undergoes a continuous transition to superradiant phase (SR), characterized by,
\begin{eqnarray}
	s_{z}^{\ast} &=&
	-\frac{\kappa}{2f_c}+\sqrt{\frac{\kappa}{f_c}}\left(\frac{\omega-\omega_0-
		1}{\sqrt{2\lambda^2-f_c\kappa}}\right),\quad
	n^{\ast} =
	\frac{f_c}{\kappa}(1-s_{z}^{*2}),\,\,\,\sin(\phi^*-\psi^*) =
	-\frac{\sqrt{f_c\kappa}}{\lambda \sqrt{2}},
	\label{supdiss}
\end{eqnarray}
where $\phi^*$ and $\psi^*$ are phases of spin and photon field respectively. The critical coupling strength corresponding to the dissipative transition between NP and SR phases is given by,
\begin{eqnarray}
\lambda_c(\kappa) = \sqrt{\frac{\kappa f_c}{2}}\sqrt{1+\frac{4(\omega_0-\omega+ 1)^2}{(2f_c-\kappa)^2}}.
\end{eqnarray}
As a result of U(1) symmetry, the phase $\phi+\psi$ increases linearly in time, however, the photon number and spin polarization remain at the steady value $n^*, s_{z}^*$ (see Fig.\ref{fig1}(a)). To describe the SR phase as a steady state, it is necessary to switch into a rotating frame with the frequency $\Omega$, where the photon field and spin projection in $x-y$ plane evolve as $\alpha_i \rightarrow \alpha_i \exp(\iota\Omega t)$ and $s^+_i\rightarrow s^+_i \exp(-\iota\Omega t)$ \cite{Pelstar}. From the EOM, the frequency $\Omega$ of the rotating frame is given by,
\begin{eqnarray}
	\Omega &=&
	\omega-1-\frac{1}{2}\sqrt{\frac{\kappa}{f_c}}\,\sqrt{2\lambda^2-\kappa
		f_c},
\end{eqnarray}
which eliminates the time evolution of $\phi+\psi$, so that it acquires a fixed arbitrary value. As a consequence of this U(1) symmetry (arbitrary $\phi+\psi$), the SR phase forms a continuous FPs lying on a circle in $x$-$p$ and $s_{x}$-$s_{y}$ plane with corresponding radius $\sqrt{2n^*}$ and $\sqrt{1-s_z^{*2}}$ respectively. 

The SR phase becomes unstable above a certain coupling strength $\lambda_{\rm I}(\kappa)$, and the system exhibits oscillatory dynamics (see Fig.\ref{fig1}(b)), which can be diagnosed by a single frequency peak in the fast Fourier transform (FFT) of $n_i(t),s_{zi}(t)$, as shown in the inset of Fig.\ref{fig1}(b). 
With increasing $\lambda$, an intermediate regime emerges, where the dynamics exhibits oscillatory or chaotic behavior depending on the initial condition.
The FFT of the chaotic trajectory shows the spreading of frequencies, whereas, for oscillatory dynamics, the FFT exhibits a single frequency.
Further increasing the coupling strength, a chaotic phase is identified, as shown in the phase diagram, given in Fig.1(a) of the main text, where trajectories display a broadened power spectrum, as depicted in the inset of Fig.\ref{fig1}(c).
Chaotic dynamics can also be detected using the Lyapunov exponent (LE), which measures the sensitivity of the initial condition. Typically, for a chaotic trajectory, a small initial perturbation $\delta \mathbf{X}(t=0)$ at the phase space point $\mathbf{X}$ grows exponentially in time $||\delta\mathbf{X}(t)||=e^{\Lambda t} \,||\delta\mathbf{X}(0)||$, which yields the Lyapunov exponent \cite{Strogatz} as,
\begin{eqnarray}
	\Lambda&=&\lim_{t \to \infty}\frac{1}{t}\ln \left(\frac{||\delta\mathbf{X}(t)||}{||\delta\mathbf{X}(0)||}\right).
	\label{LE}
\end{eqnarray}
We compute the mean Lyapunov exponent (LE) $\bar{\Lambda}$, averaged over an ensemble of initial phase space points to quantify the degree of chaoticity, as shown in Fig.1(a) of the main text. In the oscillatory phase (OP), the LE is vanishingly small, on the contrary, chaotic regime is identified by large LE, as evident from Fig.1(a) of the main text.

To analyze the state of the system in the dynamically unstable regime with no stable FPs, we perform the truncated Wigner approximation (TWA), where the initial phase space points are sampled from a Gaussian distribution to mimic the quantum fluctuation. Each initial states are evolved separately up to a sufficiently long time and finally, we obtain the ensemble average of any observable as $\langle O\rangle_{\scriptscriptstyle\text{TWA}} = \sum_j O_j(t)/{\rm N}$, where N is the number of trajectories with index $j$. 
Both in the oscillator and chaotic regime, even though the observables $O_j(t)$ along the individual classical trajectories exhibit time-dependent (irregular) oscillations, their ensemble at long time attains a stationary distribution (see Fig.\ref{fig1}(e,f)), with time-independent ensemble average $\langle O\rangle_{\scriptscriptstyle\text{TWA}}$. 
The appearance  of such stationary distribution of the observables describes the emergence of a steady state in the dynamically unstable regime,  which smoothly connects to the stable SR phase. The evolution of the photon distribution in three different dynamical regimes is shown in Fig.\ref{fig1}(d-f), which clearly manifests the delta-like distribution at $n^*$ in the SR phase as it corresponds to the stable fixed point.
 

\section{Quantum trajectory and steady state properties}
\begin{figure}[h]
	\includegraphics[width=0.8\linewidth]{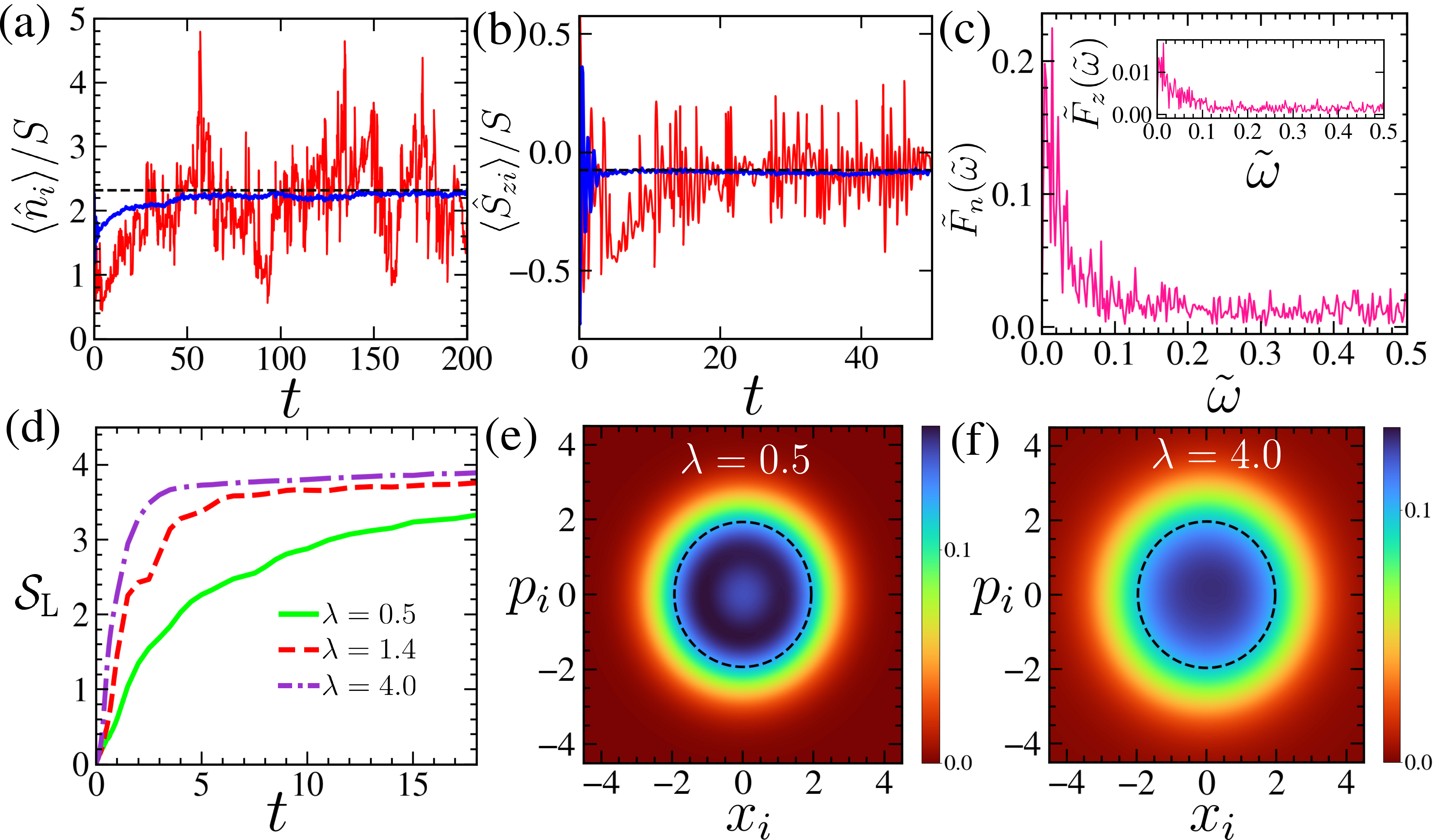}
	\caption{ {\it Quantum ergodicity:} Time evolution of (a) photon number $\langle \hat{n}_i\rangle$ and (b) spin polarization $\langle \hat{S}_{zi}\rangle$ of one of the cavities along single quantum trajectory (red) and their ensemble average (blue) in the chaotic regime. Black dashed lines are time average value of the single trajectory. (c) Fourier spectrum of $\langle \hat{n}_i\rangle$ (inset corresponds to $\langle \hat{S}_{zi}\rangle$) along single quantum trajectory. (d) The evolution of entanglement entropy $\mathcal{S}_{\rm L}$ of the left cavity for different $\lambda$. The Husimi distribution of the photon field in $x-p$ plane of one cavity in the (e) stable SR phase ($\lambda=0.5$) and (f) chaotic phase ($\lambda=4.0$). The black dashed circles in (e, f) represent the FPs of stable and unstable SR phase respectively.
		Panels (a-c) correspond to $\lambda=4.0$.
		We set $\hbar,k_B = 1$ and $\omega=2.0,\omega_0=0.5, \kappa = 0.05, \gamma_{\uparrow}=0.2,\gamma_{\downarrow}=0.1, S=2$ for all figures unless otherwise mentioned. }
	\label{fig2}
\end{figure}

Here, we consider the nonequilibrium dynamics of the open TCD, particularly emphasizing the properties of the individual quantum trajectories, within the stochastic wavefunction approach. 
To investigate the ergodic property,  we consider the time evolution of the observables along a typical quantum trajectory and obtain the time average of the observables such as $\langle \hat{n}_i\rangle_t$ and $\langle \hat{S}_{zi}\rangle_t$ up to sufficiently long time. Quantum mechanically, the ensemble average of the observables are described by their steady-state values  $\langle \hat{n}_i\rangle_{\rm ss}$ and $\langle \hat{S}_{zi}\rangle_{\rm ss}$ (independent of the initial conditions), obtained from the steady state density matrix $\hat{\rho}_{\rm ss}$. 
As evident from Fig.\ref{fig2}(a,b), $\langle \hat{n}_i\rangle_t$ and $\langle \hat{S}_{zi}\rangle_t$ approach to $\langle \hat{n}_i\rangle_{\rm ss}$ and $\langle \hat{S}_{zi}\rangle_{\rm ss}$ respectively, clearly manifesting the ergodic nature of the steady state. Similar to the semiclassical analysis, the steady state remains ergodic across all dynamical phases over the large range of coupling strength $\lambda$. 
Additionally, in the chaotic regime, the Fourier spectrum of a typical quantum trajectory spreads over a range of frequencies, as depicted in Fig.\ref{fig2}(c), indicating a signature of chaos \cite{T_Geisel}. 

To diagnose chaotic mixing, we study the nonequilibrium dynamics, starting from an arbitrary pure state, and obtain the evolution of entanglement entropy (EE) corresponding to one of the cavities. Both the growth rate and saturation value of EE increase in approach to the chaotic regime with increasing $\lambda$ (see Fig.\ref{fig2}(d)).


To understand the state of the photon field in the steady state of different regimes, we compute the Husimi distribution $Q(\alpha,\alpha^*)=\frac{1}{\pi}\bra{\alpha}\hat{\rho}_{\rm Ph}\ket{\alpha}$ from the reduced density matrix $\hat{\rho}_{\rm Ph}$ of the photon field in one of the cavities, describing the semiclassical phase space density in $x-p$ plane. 
In the stable SR phase, the Husimi distribution forms a ring-like structure, peaked near the circle of classical fixed points with a radius $\sqrt{2n^*}$ and its spreading occurs as a result of quantum fluctuations (see Fig.\ref{fig2}(e)).
In contrast, the semiclassical phase space density of photons in the chaotic regime is peaked at the center with an extended tail, resembling the thermal distribution, as shown in Fig.\ref{fig2}(f).

\begin{figure}[h]
	\includegraphics[width=0.6\linewidth]{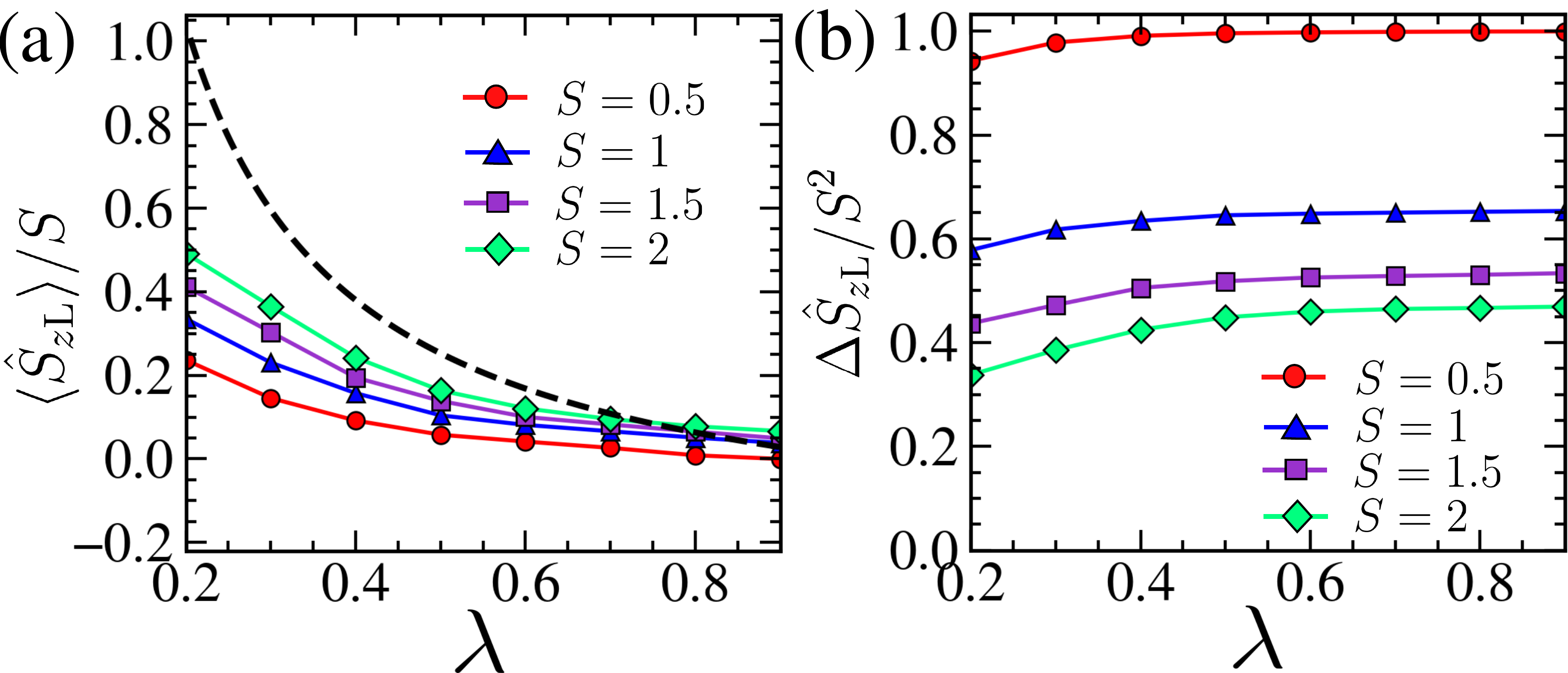}
	\caption{ {\it Approach of the results to classical limit:} (a) The steady-state value of spin polarization $\langle \hat{S}_{z\rm L}\rangle$ and (b) its fluctuation $\Delta \hat{S}_{z\rm L}$ as a function of $\lambda$ for different values of $S$. The black dashed line in panel (a) represents classical fixed point values of SR phase. We set $\omega=2.0,\omega_0=0.5, \kappa = 0.05, \gamma_{\uparrow}=0.2,\gamma_{\downarrow}=0.1$ for all panels. }
	\label{fig3}
\end{figure}

As shown in Fig.\ref{fig3}(a), we have calculated the variation of average scaled spin polarization $\langle \hat{S}_{z\rm L}\rangle/S$ for different values of $S$, which approaches the classical results with increasing $S$. Moreover, the fluctuation of the scaled spin variable $\Delta\hat{S}_{z\rm L}/S^2=(\langle \hat{S}_{z\rm L}^2\rangle-\langle \hat{S}_{z\rm L}\rangle^2)/S^2$ suppresses with increasing value of $S$ (see Fig.\ref{fig3}(b)), supporting the approach to the classical limit for large spin.

\section{Subsystem thermalization}
\begin{figure}[h]
	\includegraphics[width=0.8\linewidth]{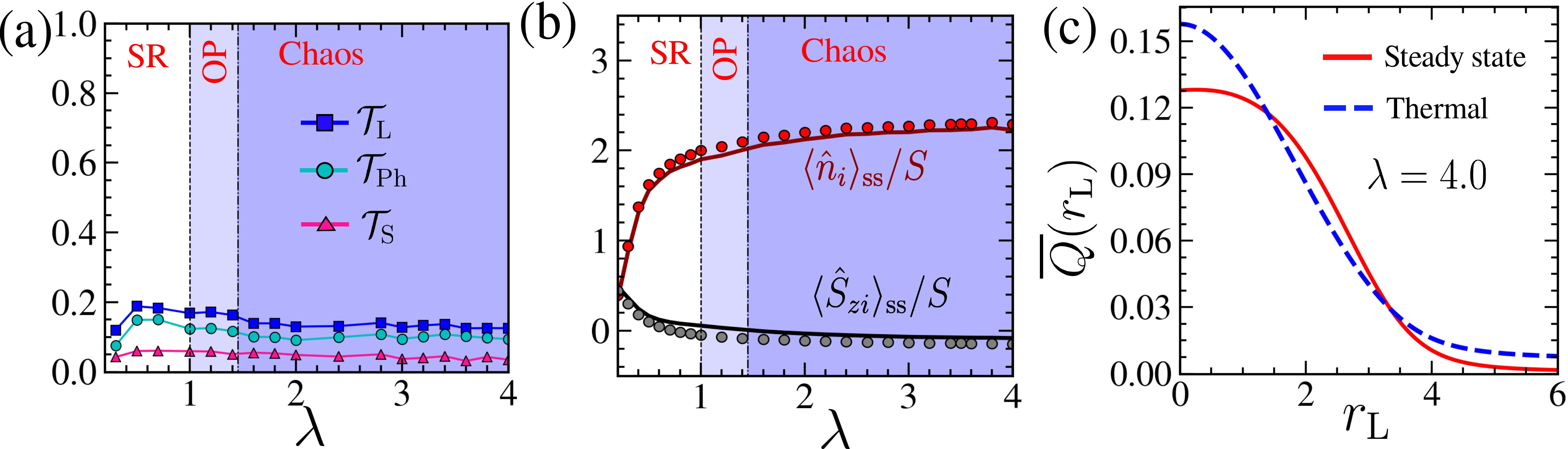}
	\caption{ Variation of (a) trace-distance $\mathcal{T}_A$ between the reduced DM of steady state $\hat{\rho}_{A}^{\rm ss}$ and thermal state $\hat{\rho}_{A}^{\rm th}$ corresponding to the left cavity ($\mathcal{T}_{\rm L}$), photon field ($\mathcal{T}_{\rm Ph}$) and spin ($\mathcal{T}_{\rm S}$) with $\lambda$. (b) The photon number $\langle \hat{n_i}\rangle_{\rm ss}$ and spin polarization $\langle \hat{S_{zi}}\rangle_{\rm ss}$ corresponding to the steady state (solid line) and thermal state (circles), as a function of $\lambda$. (c) Comparison of angular averaged Husimi $\bar{Q}(r_{\rm L})$ between the steady state (solid line) and thermal state (dashed line) in the chaotic regime ($\lambda=4.0$).  }
	\label{fig4}
\end{figure}

As discussed in the main text, the reduced density matrix (DM) of subsystem $A$, $\hat{\rho}_{A}^{\rm ss}$ corresponding to the steady state can be well described by the reduced DM $\hat{\rho}_{A}^{\rm th}$ obtained from the thermal state of the isolated TCD within generalized Gibb’s ensemble. 
We compute the trace-distance (TD) $\mathcal{T}_A=\frac{1}{2}{\rm Tr}\sqrt{(\hat{\rho}_{A}^{\rm ss}-\hat{\rho}_{A}^{\rm th})^{\dagger}(\hat{\rho}_{A}^{\rm ss}-\hat{\rho}_{A}^{\rm th})}$ between 
$\hat{\rho}_{A}^{\rm ss}$ and thermal DM $\hat{\rho}_{A}^{\rm th}$ to quantify their difference. As evident from Fig.\ref{fig4}(a), the TD remains small compared to unity over a large range of $\lambda$, indicating the similarity between $\hat{\rho}_{A}^{\rm ss}$ and $\hat{\rho}_{A}^{\rm th}$, ensuring the validity of subsystem thermalization. Consequently, the average of local observables, such as the steady state value of the photon number $\langle \hat{n}_i\rangle_{\rm ss}$ and spin polarization $\langle \hat{S}_{zi}\rangle_{\rm ss}$ are in good agreement with the corresponding values obtained from the thermal state $\hat{\rho}^{\rm th}$ (see Fig.\ref{fig4}(b)). 
In the chaotic regime, a comparison of the angular averaged Husimi distribution $\bar{Q}(r) = \int_{0}^{2\pi}Q(r,\theta) d\theta$ between the photon field of steady state and thermal state is presented in Fig.\ref{fig4}(c).

\section{Statistics of Liouvillian spectrum}
In this section, we investigate the spectral statistics of the Liouvillian superoperator of the dissipative TCD, which is given by,
\begin{eqnarray}
	\hat{\hat{\mathcal{L}}} = -i[\hat{\mathcal{H}}\otimes\mathbb{I}-\mathbb{I}\otimes\hat{\mathcal{H}}]
	+ \kappa \sum_i\mathcal{D}[\hat{a}_i]  +\frac{1}{S}\sum_i(\gamma_{\uparrow} \mathcal{D}[\hat{S}_{i+}]+
	\gamma_{\downarrow}
	\mathcal{D}[\hat{S}_{i-}]),
\end{eqnarray} 
with $\mathcal{D}[\hat{L}] = \frac{1}{2} \left(2\hat{L}\otimes \hat{L}^*-\hat{L}^{\dagger}\hat{L}\otimes \mathbb{I}-\mathbb{I}\otimes \hat{L}^{\rm TR}\hat{L}^*\right)$, where the superscript TR denotes the transposition. For numerical computation, we construct the Liouvillian $\hat{\hat{\mathcal{L}}}$ given above considering the spin magnitude $S=2$ and photon number cutoff $N_c$.
%
%
Here, we focus on the spectral properties of the Liouvillian by appropriately choosing the parameter regime, where the classical dynamics is maximally chaotic, as quantified by the Lyapunov exponent (shown in the phase diagram given in Fig.1(a) of the main text).
As mentioned in the main text, the isolated TCD system without dissipation has a U(1) symmetry, leading to the conservation of total excitation number $\hat{\mathcal{N}} = \sum_i
	(\hat{a}_i^{\dagger}\hat{a}_i+S+\hat{S}_{zi})$. 
Although the excitation number is not conserved in the open TCD, the Liouvillian $\hat{\hat{\mathcal{L}}}$ exhibits a weak symmetry associated with the superoperator $\hat{\hat{\mathcal{N}}}_s=\hat{\mathcal{N}}\otimes \mathbb{I}-\mathbb{I}\otimes\hat{\mathcal{N}}$, which satisfies $[\hat{\hat{\mathcal{L}}}, \hat{\hat{\mathcal{N}}}_s] = 0$. Consequently, the Liouvillian $\hat{\hat{\mathcal{L}}}$ can be expressed in the block diagonal form, with each block corresponding to the eigenvalues $|\mathcal{N}_s|  = 0,1,2,..$ of operator $\hat{\hat{\mathcal{N}}}_s$.
Moreover, the Hamiltonian of the TCD remains invariant under the exchange of two cavities ${\rm L} \leftrightarrow {\rm R}$ indicating the left-right symmetry of the system and the corresponding symmetry operator $\hat{\mathbb{O}}_{\rm ex}$ has eigenvalues $\pm1$. The Liouvillian also possesses a discrete symmetry corresponding to the exchange superoperator $\hat{\hat{\mathbb{O}}}_s=\hat{\mathbb{O}}_{\rm ex}\otimes \hat{\mathbb{O}}_{\rm ex}$, satisfying $[\hat{\hat{\mathcal{L}}}, \hat{\hat{\mathbb{O}}}_s] = 0$.
To perform the spectral analysis of the non-hermitian superoperator $\hat{\hat{\mathcal{L}}}$, we consider its eigenvalues corresponding to a fixed symmetry sector of both $\hat{\hat{\mathcal{N}}}_s$ and $\hat{\hat{\mathbb{O}}}_s$. 
All the results presented below remain unchanged regardless of whether we consider the even or odd sector of exchange symmetry. Here, we present the results for the even sector (+1 eigenvalue of $\hat{\hat{\mathbb{O}}}_s$).

\begin{figure}[t]
	\includegraphics[width=0.9\linewidth]{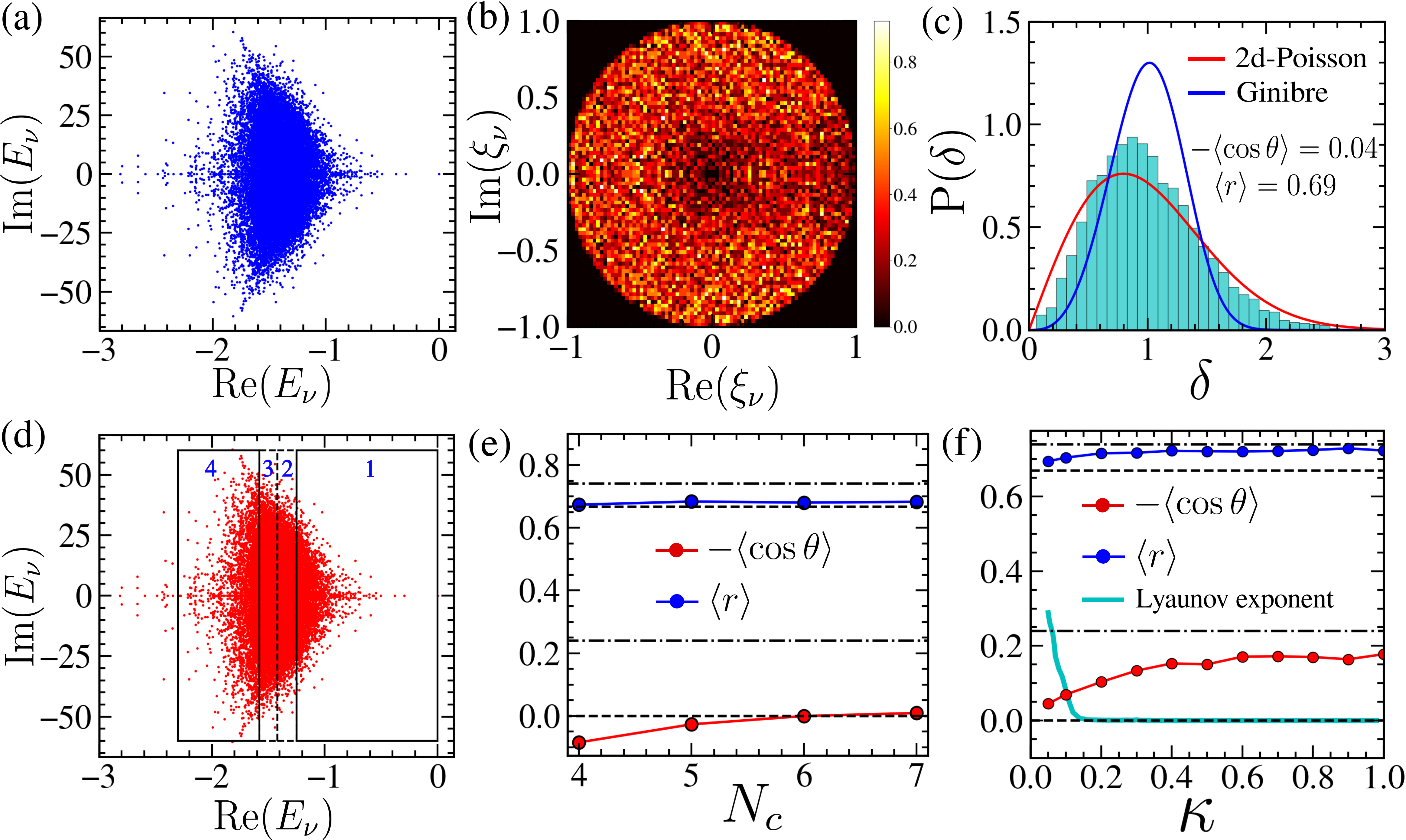}
	\caption{ {\it Statistics of the Liouvillian spectrum in the classically chaotic regime:} (a) The eigenvalues of the Liouvillian corresponding to the block with $\mathcal{N}_s = 0$ in the complex plane. (b) The corresponding distribution of the CSR $\xi_{\nu}$ as color scale in the complex plane  with the characteristic measures -$\langle \cos\theta\rangle = 0.04,\,\langle r\rangle = 0.69$. (c) The level spacing distribution P($\delta$). 
	(d) The spectrum in the panel (a) divided into four different bands (marked by numbers)	along real axis.
	(e) The variation of $\langle r\rangle$ and -$\langle \cos\theta\rangle$ with cutoff $N_{\rm c}$ for the block $\mathcal{N}_s = 6$. 
	Panels (a-e) are obtained for $\lambda = 4.0,\kappa=0.05$, where classical dynamics is strongly chaotic (see phase diagram in Fig.1(a) of main text). (f) Variation of $\langle r\rangle$, -$\langle \cos\theta\rangle$, average Lyapunov exponent with photon decay rate $\kappa$ for coupling strength $\lambda=4$.
	Here, dashed (dashed-dotted) line represents the corresponding value of 2d-Poisson (Ginibre) statistics. The panels (a,b,c,d,f) are obtained for number of photonic states $N_c=5$ and for the block with $\mathcal{N}_s = 0$. The panel (e) is obtained from the full spectrum for the block with $\mathcal{N}_s = 6$. All the results are obtained using the eigenvalues with positive sector of exchange symmetry.
	Parameter chosen: $\gamma_{\downarrow}=0.1,\gamma_{\uparrow}=0.2$. }
	\label{fig5}
\end{figure}

The pattern of the complex eigenvalues $E_{\nu}$ of the full spectrum of the truncated Liouvillian with cutoff $N_c=5$, belonging to $\mathcal{N}_s=0$ and even exchange symmetry sector is shown in Fig.\ref{fig5}(a), which is symmetric about the negative real axis.
Next, we compute the complex spacing ratio (CSR) \cite{Prosen_PRX},
\begin{eqnarray}
	\xi_{\nu} = \frac{E_{\nu}^{\rm NN}-E_{\nu}}{E_{\nu}^{\rm NNN}-E_{\nu}} = r_{\nu} e^{i\theta_{\nu}},
\end{eqnarray}
where $E_{\nu}^{\rm NN}$ and $E_{\nu}^{\rm NNN}$ are the nearest and next-nearest neighbors of the complex eigenvalue $E_{\nu}$, which is a generalization of the adjacent gap ratio in the Hamiltonian system. It is evident from Fig.\ref{fig5}(b) that the
CSR  $\xi_{\nu}$ is distributed uniformly within a circle in the complex plane, reflecting the universal feature of independent complex numbers, which follows 2d-Poisson statistics of random matrix theory (RMT) \cite{Prosen_PRX}. To quantify such universality in a spectrum, we compute $\langle r\rangle$ and $\langle \cos\theta\rangle$, which agrees well with that of 2d-Poisson distribution with $\langle r\rangle\approx0.67$ and $\langle \cos\theta\rangle = 0$.
From the raw data of level spacing (Euclidean distance $\delta$ between the nearest neighbors), we perform the unfolding procedure (following the method as described in \cite{Prosen_PRL, Kulkarni_Dicke}) to compute level spacing distribution, which also agrees with the 2d-Poisson distribution, given by ${\rm P}_{\rm 2d-P}(\delta) = \frac{\pi}{2}\delta \exp(-\pi\delta^2/4)$ \cite{Prosen_PRX}, as depicted in Fig.\ref{fig5}(c). 
%
%
Here we point out that the level repulsion in the spectrum is manifested by the level spacing distribution following the GinUE statistics of RMT with P$(\delta)\sim\delta^3$ for small $\delta$. The corresponding CSR exhibits $\langle r\rangle\approx0.74$ and -$\langle \cos\theta\rangle=0.24$ \cite{Prosen_PRX}. However,  in the present case, we do not find any signature of level repulsion in the Liouvillian spectrum corresponding to the classical chaotic regime. 

To unveil any hidden Ginibre correlation within certain parts of the Liouvillian spectrum arising from dissipation, we divide the eigenvalues corresponding to a specific symmetry sector into distinct bands along the negative real axis and analyze the spectral statistics of each band separately, as shown in Fig.\ref{fig5}(d). For this purpose, it is preferable to analyze the spectrum corresponding to the largest block with $\mathcal{N}_s=0$, which contains the zero eigenvalue and spreads along the real axis. 
Within our computational capacity, we can diagonalize this $\mathcal{N}_s = 0$ block (as well as the complete Liouvillian) for only up to photon number cutoff $N_{\rm c}=5$. As depicted in Fig.\ref{fig5}(d), the bands at the edges are of larger width, so that each region contains similar number of eigenvalues, sufficient for statistical analysis. Additionally, in a similar manner, we have also analyzed the spectral statistics for a few large blocks with $\mathcal{N}_s= 1,2$.	
%

%
The values of the quantities $\langle r\rangle$ and $\langle \cos\theta\rangle$ for the spectrum in each band, corresponding to different blocks of $\hat{\hat{\mathcal{N}}}_s$ (with even exchange symmetry) are presented in Table.\ref{tab:Strips_statistics}. It is evident that there is no hidden Ginibre correlation in any part of the spectrum and the statistics of the respective eigenvalues remains close to the 2d-Poisson distribution.

\begin{table}[h]
	\centering
	\label{tab:Strip_regions}
	\begin{tabular}{|c|c|c|c|c|}
		\hline
		\textbf{$\boldsymbol{\mathcal{N}_s}$} &\quad \textbf{Band 1} \quad&\quad \textbf{Band 2}\quad & \quad\textbf{Band 3} \quad&\quad \textbf{Band 4}\quad \\
		\hline
		
		0  & -$\langle \cos\theta \rangle =$ -0.016, $\langle r\rangle = 0.676$  & -$\langle \cos\theta \rangle =$ 0.004, $\langle r\rangle = 0.680$  & -$\langle \cos\theta \rangle =$ -0.001, $\langle r\rangle = 0.690$  & -$\langle \cos\theta \rangle =$ -0.033, $\langle r\rangle = 0.677$  \\
		1  & -$\langle \cos\theta \rangle =$ -0.020, $\langle r\rangle = 0.683$  & -$\langle \cos\theta \rangle =$ -0.032, $\langle r\rangle = 0.681$  & -$\langle \cos\theta \rangle =$ -0.014, $\langle r\rangle = 0.681$  & -$\langle \cos\theta \rangle =$ -0.040, $\langle r\rangle = 0.678$  \\
		2  & -$\langle \cos\theta \rangle =$ -0.037, $\langle r\rangle = 0.670$   & -$\langle \cos\theta \rangle =$ -0.044, $\langle r\rangle = 0.676$   & -$\langle \cos\theta \rangle =$ -0.006, $\langle r\rangle = 0.689$ & -$\langle \cos\theta \rangle =$ -0.055, $\langle r\rangle = 0.673$  \\
		\hline
	\end{tabular}
	\caption{Spectral statistics in terms of -$\langle \cos\theta \rangle$ and $\langle r \rangle$ of the full spectrum corresponding to the photon cutoff $N_c = 5$ and different block with $\mathcal{N}_s$ and even sector of exchange symmetry for each of the four bands.}
	\label{tab:Strips_statistics}
\end{table}

Next, we investigate the variation of statistics with increasing photon cutoff $N_c$.
For this purpose, we obtain the spectrum of $\hat{\hat{\mathcal{L}}}$ for a relatively smaller block, corresponding to the eigenvalue $\mathcal{N}_s=6$, which can be diagonalized up to the photon cutoff $N_c=7$. We obtain the quantities $\langle r\rangle$ and $\langle \cos\theta\rangle$ of the eigenvalues belonging to the positive exchange symmetry with increasing $N_c$, as summarized in Table.\ref{tab:eigenvalues} and also shown in Fig.\ref{fig5}(e). Clearly, these quantities $\langle r\rangle$ and $\langle \cos\theta\rangle$ converge to that of 2d-Poisson distribution.

\begin{table}[h]
	\centering
	\begin{tabular}{|c|c|c|c|c|}
		\hline
		\textbf{Cutoff} ($\boldsymbol{N_c}$) & \textbf{Total Eigenvalues} &-$\boldsymbol{\langle \cos\theta \rangle}$ & $\boldsymbol{\langle r \rangle}$ \\
		\hline
		
		4   & 2304   & \quad -0.08429 \quad &\quad 0.673534 \quad \\
		5  & 6544   & \quad -0.02664 \quad & \quad 0.683568 \quad \\
		6  & 15036    & \quad -0.000129 \quad& \quad 0.680234 \quad \\
		7  & 29677    & \quad 0.009490 \quad& \quad 0.6826832 \quad \\
		\hline
	\end{tabular}
	\caption{Variation of -$\langle \cos\theta \rangle$ and $\langle r \rangle$ with photon cutoff $N_c$ for the full spectrum of the block with $\mathcal{N}_s = 6$ and positive exchange symmetry sector.}
	\label{tab:eigenvalues}
\end{table}

Furthermore, we analyze the spectral statistics using only the set of converged eigenvalues. To identify this converged spectrum, we select those eigenvalues (within a small relative error), which do not change appreciably as the photon cutoff is increased from $N_c$ to $N_c+1$, satisfying the following criteria,
%
\begin{eqnarray}
	\frac{|E_{\nu}^{\rm Nc}-E_{\nu }^{\rm Nc+1}|}{|E_{\nu}^{\rm Nc}|} \le 5\cross10^{-3},
\end{eqnarray}
where $E_{\nu}^{\rm Nc}$ are eigenvalues corresponding to photon truncation $N_c$. We have checked that the absolute error also remains small $\Delta E_{\nu} = |E_{\nu}^{\rm Nc}-E_{\nu }^{\rm Nc+1}| < 10^{-2}$.
%
%
Next, we consider a relatively smaller block with $\mathcal{N}_s = 6$ and positive sector of exchange symmetry to calculate the fraction of eigenvalues that are converged as the photon cutoff increases and also analyze their statistical properties.
%
%
The values of $\langle r\rangle$ and $\langle \cos\theta\rangle$  of these converged eigenvalues for different cutoff $N_c$ are given in Table.\ref{tab:Converged_eigenvalues}, which also remains closer to the 2d Poisson distribution. 
%
%

\begin{table}[h]
	\centering
	\begin{tabular}{|c|c|c|c|c|}
		\hline
		\textbf{Cutoff}($\boldsymbol{N_c\rightarrow N_c+1}$) & \textbf{Total Eigenvalues with $\boldsymbol{N_c}$} & \textbf{Converged Eigenvalues} & -$\boldsymbol{\langle \cos\theta \rangle}$\quad & $\boldsymbol{\langle r \rangle}$\quad \\
		\hline
		
		4 $\rightarrow$ 5  & 2304  & 1558 (62.62 \% converged)  &\quad -0.1052 \quad & \quad 0.67399 \quad \\
		5 $\rightarrow$ 6  & 6544  & 5219 (79.75 \% converged) & \quad-0.03439 \quad & \quad 0.67627 \quad \\
		6 $\rightarrow$ 7  & 15036  & 13130 (87.32 \% converged)  &\quad -0.01157 \quad& \quad 0.67843 \quad \\
		\hline
	\end{tabular}
\caption{Spectral statistics of the set of converged eigenvalues for photon cutoff $N_c$ (converged by increasing photon cutoff from $N_c$ to $N_c+1$) in terms of -$\langle \cos\theta \rangle$ and $\langle r \rangle$ corresponding to the block with $\mathcal{N}_s = 6$ and positive exchange symmetry sector. The fraction of converged eigenvalues increases with photon cutoff $N_c$.}
\label{tab:Converged_eigenvalues}
\end{table}

Additionally, similar to the previous analysis, we have also investigated the statistical properties of the converged eigenvalues by dividing them into different bands to uncover any hidden Ginibre correlation. For this purpose, we obtain the converged eigenvalues for $N_c = 4$ (converged as cutoff changes from $N_c = 4$ to 5) corresponding to a few large blocks with $\mathcal{N}_s = 0,1,2$ and positive sector of exchange symmetry.
%
%
Due to the insufficient number of converged eigenvalues, we divide the spectrum only into two bands ${\Re}(E_{\nu}) \ge -1.3$ and ${\Re}(E_{\nu}) < -1.3$ along the real axis (each contains $\sim 3000$ eigenvalues ).
The values of the quantities -$\langle \cos\theta \rangle$ and $\langle r \rangle$ for the spectrum in each band, corresponding to different blocks of $\hat{\hat{\mathcal{N}}}_s$ are presented in Table.\ref{tab:Strips_of_converged_eigenvalues}, which shows a close agreement to that of 2d-Poisson distribution.


\begin{table}[H]
	\centering
	\label{tab:Strip_regions_Converged}
	\begin{tabular}{|c|c|c|c|c|}
		\hline
		\textbf{$\boldsymbol{\mathcal{N}_s}$} &\quad \textbf{Band 1 $ (  \boldsymbol{{\Re}(E_{\nu}) \ge -1.3})$}  \quad&\quad \textbf{Band 1 $ (  \boldsymbol{{\Re}(E_{\nu}) < -1.3})$}\quad \\
		\hline

		0  & -$\langle \cos\theta \rangle =$ -0.0442, $\langle r\rangle = 0.67520$  & -$\langle \cos\theta \rangle =$ -0.0450, $\langle r\rangle = 0.67922$   \\
		1  & -$\langle \cos\theta \rangle =$ -0.0654, $\langle r\rangle = 0.67513$  & -$\langle \cos\theta \rangle =$ -0.0479, $\langle r\rangle = 0.68255$ \\
		2  & -$\langle \cos\theta \rangle =$ -0.0462, $\langle r\rangle = 0.67662$  & -$\langle \cos\theta \rangle =$ -0.0394, $\langle r\rangle = 0.67274$   \\
		\hline
	\end{tabular}
\caption{Spectral statistics of the converged spectrum in terms of -$\langle \cos\theta \rangle$ and $\langle r \rangle$ for the photon cutoff $N_c = 4$ (converged by increasing photon cutoff from $N_c = 4$ to $5$). The spectrum is obtained for different block of $\mathcal{N}_s$ and even sector of exchange symmetry for two distinct bands.}
\label{tab:Strips_of_converged_eigenvalues}
\end{table}

%

However, for increasing the dissipation strength $\kappa$, we find that the Liouvillian spectrum exhibits a tendency towards Ginibre correlation. To demonstrate this, we perform the spectral analysis of the full spectrum of the Liouvillian with $N_c = 5$ (considering the largest block with $\mathcal{N}_s = 0$ and +1 eigenvalue of exchange symmetry). As shown in Fig.\ref{fig5}(f), both the $-\langle \cos\theta\rangle$ and $\langle r\rangle$ approach to the limit corresponding to the Ginibre distribution for sufficiently large $\kappa$. In this regard, we emphasize that the chaos is suppressed for larger values of dissipation strength ($\kappa \gtrsim 0.1$, which is evident from the decay of the average Lyapunov exponent, as shown in Fig.\ref{fig5}(f), as well as from Fig.1(a) of the main text), even though the spectral statistics exhibits a Ginibre correlation. According to the GHS conjecture  \cite{Haake_GHS}, the Ginibre correlation in the spectral statistics of Liouvillian is considered to be the signature of chaos in dissipative systems.
%
Recently, such violation of GHS conjecture in an atom-photon system has been reported in \cite{Lea_Santos_GHS}, where the spectral statistics of Liouvillian exhibits Ginibre correlation in both regular and chaotic regimes. The present analysis indicates a stronger violation of the GHS conjecture as the spectral statistics exhibits a crossover from 2d-Poisson to Ginibre distribution, while the corresponding classical dynamics changes from chaotic to regular behavior as dissipation strength $\kappa$ increases (see Fig.\ref{fig5}(f)).

Ginibre distribution can also be observed in a few similar models, such as the anisotropic Dicke model with photon loss \cite{Lea_Santos_GHS,Kulkarni_Dicke} as well as in Tavis-Cummings dimer in the presence of both gain and loss in the two cavities \cite{TCD_Tomoghno}. The Hamiltonian for the anisotropic Dicke model is given by,
%
%
	\begin{eqnarray}
		\hat{\mathcal{H}}_{\rm AD} = \omega\, \hat{a}^{\dagger}\hat{a}+\omega_0\, \hat{S}_{z}+\frac{\lambda_-}{\sqrt{2S}}(\hat{a}\hat{S}_{+}+\hat{a}^{\dagger}\hat{S}_{-})+\frac{\lambda_+}{\sqrt{2S}}(\hat{a}\hat{S}_{-}+\hat{a}^{\dagger}\hat{S}_{+}),
	\end{eqnarray}
where $\lambda_-$ and $\lambda_+$ are the atom-photon coupling strengths. 

In references \cite{Lea_Santos_GHS,Kulkarni_Dicke}, this system is considered in the presence of photon loss with rate $\kappa = 1$.
In order to investigate the spectral properties of the Liouvillian, we numerically construct the matrix for spin magnitude $S = 3$ and photon cutoff $N_c = 30$. Since the Liouvillian possesses the weak parity symmetry, we choose its positive sector of the full spectrum. First, we consider the system in the Dicke limit $\lambda_+ = \lambda_-$ and for the coupling parameters $\lambda_+ = \lambda_- = 1,$ $\kappa = 1$, we obtain $\langle r\rangle \approx 0.729$ and -$\langle \cos\theta\rangle \approx 0.229$ from the level spacing ratio (the rest of the parameters such as photon frequency $\omega = 1$ and energy gap of two-level system $\omega_0=1$ are also chosen to be the same as that of \cite{Lea_Santos_GHS,Kulkarni_Dicke}). This corresponds to the Ginibre distribution, which is in agreement with the results of \cite{Lea_Santos_GHS,Kulkarni_Dicke}. 
%
%

Furthermore, away from the Dicke limit with $\lambda_- = 1,\,\lambda_+ = 2$ and photon loss rate $\kappa = 1$, the spectrum of the Liouvillian (for $S = 3$ and $N_c = 30$) corresponding to the positive parity sector yields $\langle r\rangle \approx 0.734$ and -$\langle \cos\theta\rangle \approx 0.233$ revealing the Ginibre distribution, similar to the work \cite{Lea_Santos_GHS} (other parameters are also 
chosen to be the same as that of \cite{Lea_Santos_GHS}).
%

Additionally, the Ginibre correlation has also been found in the spectral statistics of the Tavis-Cummings dimer coupled to a Markovian bath \cite{TCD_Tomoghno}, describing the gain and loss of photons in the two cavities with equal rate $\kappa$. The Hamiltonian for this system is the same as that of the present work (given in Eq.(1) of the main text). The Liouvillian is given by,
\begin{eqnarray}
	\hat{\hat{\mathcal{L}}}_{\rm TCD} = -i[\hat{\mathcal{H}}\otimes\mathbb{I}-\mathbb{I}\otimes\hat{\mathcal{H}}]
	+ \sum_{i=1,2} \left(2\hat{L}_i\otimes \hat{L}_i^*-\hat{L}_i^{\dagger}\hat{L}_i\otimes \mathbb{I}-\mathbb{I}\otimes \hat{L}_i^{\rm TR}\hat{L}_i^*\right),
\end{eqnarray} 	
which is constructed with jump operators $\hat{L}_1 = \sqrt{\kappa} \hat{a}_{\rm L}$  and $\hat{L}_2 = \sqrt{\kappa} \hat{a}_{\rm R}^{\dagger}$, describing the loss and gain of photons from the left and right cavities respectively, which breaks the exchange symmetry of the system.
%
%
This Liouvillian possesses a weak symmetry corresponding to the superoperator $\hat{\hat{\mathcal{N}}}_s$ (as mentioned previously), due to which it is decomposed into blocks and the spectral analysis is performed for the largest block with eigenvalue $\mathcal{N}_s = 0$, as performed in \cite{TCD_Tomoghno}.
Numerically, the Liouvillian is constructed for $S=2$ and $N_c = 5$. For the parameters $\lambda = 1$, $\kappa = 1$, we obtain  $\langle r\rangle \approx 0.7256$ and -$\langle \cos\theta\rangle \approx 0.2082$ indicating a Ginibre correlation, as observed in \cite{TCD_Tomoghno} (other parameters are also chosen to be the same as that of \cite{TCD_Tomoghno}).